\documentclass[journal=jceaax,manuscript=article]{achemso}

\usepackage{graphicx}\graphicspath{{./IMAGES/}}
\usepackage[version=3]{mhchem}
\usepackage{siunitx}
\DeclareSIUnit\angstrom{\text{Å}}
\DeclareSIUnit{\calorie}{cal}
\DeclareSIUnit{\bar}{bar}
\DeclareUnicodeCharacter{0308}{\\"{}}
\usepackage[hidelinks]{hyperref}
\usepackage[nameinlink,capitalize]{cleveref}
\usepackage{subcaption}
\usepackage{booktabs}
\usepackage{caption}
\usepackage{xcolor}
\usepackage{lineno}

\SectionNumbersOn
\setkeys{acs}{maxauthors = 0}
\setkeys{acs}{doi = true}
 \setkeys{acs}{keywords = true}
\author{D. Raju}
\affiliation{Engineering Thermodynamics, Process \& Energy Department, Faculty of Mechanical Engineering, Delft University of Technology, Leeghwaterstraat 39, Delft 2628CB, The Netherlands}
\author{M. Ramdin}
\affiliation{Engineering Thermodynamics, Process \& Energy Department, Faculty of Mechanical Engineering, Delft University of Technology, Leeghwaterstraat 39, Delft 2628CB, The Netherlands}
\author{T.J.H. Vlugt}
\affiliation{Engineering Thermodynamics, Process \& Energy Department, Faculty of Mechanical Engineering, Delft University of Technology, Leeghwaterstraat 39, Delft 2628CB, The Netherlands}
\email{T.J.H.Vlugt@tudelft.nl}

\title{Thermophysical Properties and Phase Behavior of \ce{CO2} with Impurities: Insight from Molecular Simulations}

\begin{document}
% \linenumbers
\clearpage
\begin{abstract}
Experimentally determining thermophysical properties for various compositions commonly found in \ce{CO2} transportation systems is extremely challenging. To overcome this challenge, we performed Monte Carlo (MC) and Molecular Dynamics (MD) simulations of \ce{CO2} rich mixtures to compute thermophysical properties such as densities, thermal expansion coefficients, isothermal compressibilities, heat capacities, Joule-Thomson coefficients, speed of sound, and viscosities at temperatures (235-313) K and pressures (20-200) bar. We computed thermophysical properties of pure \ce{CO2} and \ce{CO2} rich mixtures with \ce{N2}, \ce{Ar}, \ce{H2}, and \ce{CH4} as impurities (1-10) mol\% and showed good agreement with available Equation of State (EoS). We showed that impurities decrease the values of thermal expansion coefficients, isothermal compressibilities, heat capacities, and Joule-Thomson coefficients in the gas phase, while these values increase in the liquid and supercritical phases. In contrast, impurities increase the value of speed of sound in the gas phase and decrease it in the liquid and supercritical phases. We present an extensive data set of thermophysical properties for \ce{CO2} rich mixtures with various impurities, which will help to design the safe and efficient operation of \ce{CO2} transportation systems.
\end{abstract}


\clearpage
\section{Introduction}
Climate change is being driven mostly by \ce{CO2} emissions from the combustion of fossil fuels (oil, natural gas, and coal) for electricity production, transportation, heating, and other industrial applications \cite{ipcc2023,iea2023,mohitpour2012pipeline}. The 2023 statistical review of world energy indicates that more than 80\% of world energy consumption comes from fossil fuels \cite{iea2023,Statistical_Review_of_WorldEnergy_2023}. It is unfeasible to shift entirely to renewable energy resources on a very short time scale \cite{osman2023cost}. Carbon Capture and Storage (CCS) is the most popular and large-scale process used in industries to meet the anthropogenic \ce{CO2} emission targets \cite{boot2014carbon}. CCS is carried out in three different stages, namely, capture, transportation, and sequestration \cite{bui2018carbon}. Carbon capture and sequestration depends upon safe and economical transportation of \ce{CO2} \cite{onyebuchi2018systematic}. In the past four decades, pipelines have been used to successfully inject pure \ce{CO2} into depleted or nearly oil/gas fields for Enhanced Oil/gas Recovery (EOR) \cite{mohitpour2012pipeline}. The injection of \ce{CO2} captured from the flue gas stack is significantly different from pure \ce{CO2} injection for EOR \cite{munkejord2016co2}. The difference is the presence of impurities, since no gas separation process is 100\% efficient \cite{demetriades2016new}. Impure \ce{CO2} also differs in composition depending on the source and technology of capture \cite{serpa2011technical}. It is possible to obtain pure \ce{CO2} from impure \ce{CO2}, but this will result in additional costs and carbon footprint \cite{serpa2011technical,li2011pvtxy}. The most efficient and preferred phase of transporting \ce{CO2} in the pipeline is the dense supercritical or liquid phase \cite{onyebuchi2018systematic}. Due to the presence of impurities, especially non-condensable gases (for example, \ce{Ar}, \ce{N2}, \ce{H2}, \ce{O2}, and \ce{CH4} which have a low boiling point compared to \ce{CO2}) reduces the density of impure \ce{CO2} mixtures and is likely to introduce two-phase flow behavior \cite{brownsort20191st}. Two-phase flows during transportation lead to numerous challenges, mainly pressure surge \cite{yang2021improved}, which will sequentially lead to possible pipeline failure \cite{lund2011depressurization}. The recommended level of major impurities in \ce{CO2} rich stream for safe and efficient pipeline transportation and sequestration from different standards (NETL \cite{shirley_myles_2019}, Dynamis \cite{de2007towards}, and ISO \cite{simonsen_dennis}) and projects (Porthos \cite{porthos} and Teeside \cite{brownsort20191st}) are summarized in \Cref{prop_table,table1}. For quality standards of additional minor impurities in \ce{CO2} transportation systems, the reader is referred to the review article of Simonsen et al.\cite{simonsen_dennis}. \ce{CO2} rich stream with impurities significantly alters the thermodynamic and transport properties of \ce{CO2}, which will, in turn, impact the overall flow behavior, pipeline capacity, and operating window in \ce{CO2} pipeline systems \cite{nazeri2021impact,cresswell2016molecular}. Therefore, knowledge on thermodynamic and transport properties is indispensable to model the flow and phase behavior of impure \ce{CO2} rich mixture within the operating window for the safe design and efficient operation of \ce{CO2} transportation systems \cite{munkejord2016co2}.

\ce{CO2} captured from the stationary sources is compressed to a pressure higher than the critical pressure to avoid two-phase flows \cite{onyebuchi2018systematic}. Due to planned maintenance or failure, transient processes such as startup, shutdown, and depressurization are anticipated in \ce{CO2} transportation systems, which plausibly lead to two-phase flows \cite{onyebuchi2018systematic}. Hence, the operational window and conditions of \ce{CO2} pipeline systems span a broad range of temperatures and pressures, encompassing transient processes from the wellhead to the bottom of the well. Operating conditions vary based on the geological location and reservoir characteristics \cite{munkejord2016co2}. High temperatures are limited considering the temperature limit of the pipeline coating material ($<$ \SI{50}{\celsius}) and cooling after compression stages \cite{mohitpour2012pipeline}. The discharge pressure from the compressor to the pipeline is generally in the range of 100 to \SI{200}{\bar} \cite{mohitpour2012pipeline}. The lowest temperature and pressure limit depends on geological conditions and phase behavior of \ce{CO2} rich stream to maintain a dense phase \cite{serpa2011technical}. Therefore, the temperature and the pressure range expected in \ce{CO2} pipeline systems is assumed as \SI{-20}{\celsius} to \SI{40}{\celsius} and \SI{0}{\bar} to \SI{200}{\bar}, respectively. A priori operational conditions considered in this study incorporate conditions at which transient events such as startup, shutdown, and depressurization are anticipated to occur \cite{sacconi2020modelling}. Consequently, to ensure accurate modeling of transient processes, it is crucial to know the thermodynamic and transport properties for a wide range of temperatures and pressures expected within \ce{CO2} transportation systems.

The thermodynamic and transport properties of impure \ce{CO2} can be computed from thermodynamic models such as Equations of State (EoS) and other empirical correlations available in literature \cite{span1996new,kunz2012gerg,laesecke2017reference}. The validity of EoS predictions largely depends on the interaction parameters that are obtained by fitting Vapor-Liquid Equilibrium (VLE) data obtained from experiments and assumptions used to develop EoS \cite{ramdin2016computing,balaji2015simulating}. Most EoS models accurately predict the thermodynamic properties related to first-order derivatives of the thermodynamic potentials (Gibbs energy, Helmholtz energy, enthalpy, and internal energy), i.e., the phase equilibria \cite{shin2008estimation,lee2008estimation}. The second-order derivative properties, such as isothermal compressibility, thermal expansion coefficient, Joule–Thomson coefficient, heat capacity, and speed of sound, are not predicted accurately by majority of EoS models \cite{shin2008estimation,lee2008estimation}. These properties serve as a basis for designing and modeling pipeline transportation systems. Especially, knowledge on speed of sound is crucial in characterizing the state of the fluid in pipeline transportation systems \cite{fazelabdolabadi2010molecular,fazelabdolabadi2010prediction}. Many literature studies predict the thermodynamic and transport properties of impure \ce{CO2} using either a simple or an advanced EoS \cite{quinones2001one,vesovic1990transport}, but no general agreement has been made to use a particular EoS with specific interaction parameters for \ce{CO2} mixtures with small amount of impurities \cite{cresswell2016molecular,shin2018evaluation}.

Determination of the thermodynamic and transport properties from experiments is difficult due to low concentrations limits of impurities in \ce{CO2} transportation systems (\cref{table1}). Performing experiments for a wide range of compositions and conditions of a multi-component \ce{CO2} mixtures is costly and time-consuming \cite{wilhelmsen2012evaluation}. Molecular simulations with classical force fields are widely used to compute the thermodynamic and transport properties of multi-component gas systems \cite{rahbari2021effect,polat2023densities,polat2023transport}. Densities, viscosities, and phase equilibria calculated from molecular simulations provide reasonable and sometimes better predictions than EoS since calculations are based on accurate interaction potentials between atoms and molecules \cite{gecht2020mdbenchmark,ciccotti2022molecular,polat2023densities}. Simulations can efficiently compute heat capacities and speed of sound, both of which are essential for modeling transient phenomena like vapor collapse accurately \cite{fazelabdolabadi2010molecular,rahbari2021effect,cresswell2016molecular,bergant1999pipeline}. Using classical force field-based Monte Carlo (MC) simulations, Cresswell et al. \cite{cresswell2016molecular} computed phase equilibria and densities of binary mixtures of \ce{CO2}  with \ce{Ar}, \ce{N2}, \ce{H2}, and \ce{O2} for a range of temperatures from \SI{0}{\celsius} to \SI{50}{\celsius} and pressure up to \SI{200}{\bar}. Aimoli et al. \cite{aimoli2014force} evaluated the performance of different force fields for computing density, isothermal compressibility, thermal expansion coefficient, heat capacity at constant volume and pressure, Joule-Thomson coefficient, viscosity, and speed of sound for pure \ce{CO2} and \ce{CH4} for a range of temperatures from \SI{-20}{\celsius} to \SI{100}{\celsius} and pressure up to \SI{1000}{\bar} using Molecular Dynamics (MD) simulation. To the best of our knowledge, most molecular simulation studies are limited to binary \ce{CO2} mixtures \cite{cresswell2016molecular,aimoli2014force}. Molecular simulations dedicated to multi-component \ce{CO2} mixtures are extremely limited \cite{al2013effect,coquelet2017transport} or nonexistent.

In this work, we compute the thermodynamic and transport properties of pure and impure \ce{CO2} streams for a range of impurity levels ranging from (1-10) mol\% (which includes 12 \ce{CO2} binary mixtures with 1 mol\%, 5 mol\%, and 10 mol\% impurities which are shown in the Section S14 of the Supporting Information, and 36 multi-component \ce{CO2} mixtures with impurities $\leq$ 4 mol\% which are shown in the Section S17 and S18 of the Supporting Information) using MC and MD simulation techniques. Simulations were carried out for a range of temperatures from \SI{-20}{\celsius} to \SI{40}{\celsius} and pressure up to \SI{200}{\bar}. The main impurities, \ce{N2}, \ce{Ar}, \ce{H2}, and \ce{CH4}, are selected to investigate the effect of impurities on the thermodynamic and transport properties. A comprehensive list of chemical components, CAS numbers, and force fields of all components used in this work are shown in \cref{prop_table}. Properties calculated within the operating window include density, isothermal compressibility, thermal expansion coefficient, heat capacity at constant volume and pressure, Joule-Thomson coefficient, shear viscosity, and speed of sound. We showed that, in comparison to pure \ce{CO2} density, \ce{CO2} rich mixture with molecular weight lower than pure \ce{CO2} at a condition had lower densities. We also showed that impurities decrease the value of thermal expansion coefficients, isothermal compressibilities, heat capacities, and Joule-Thomson coefficients in the gas phase and increase the value of these properties in the liquid and supercritical phases. Conversely, impurities tend to increase speed of sound in the gas phase and decrease in the liquid and supercritical phases. Our results show that the order of influence of a particular impurity on a thermodynamic property other than density correlates with the critical temperature of that impurity.

This manuscript is organized as follows: the methodology used to compute thermodynamics and transport properties is explained in \cref{section-2}, followed by the simulation details in \cref{section-3}. In \cref{section-4}, we present the validation and the results of the computed thermodynamic and transport properties. In \cref{section-5}, our key findings are summarized.

\section{Theoretical Background}
\label{section-2}
To compute the speed of sound ($c$), one requires other properties, which include heat capacity at constant pressure ($C_\mathit{P}$), heat capacity at constant volume ($C_\mathit{V}$), and isothermal compressibility ($\beta_T$) \cite{fazelabdolabadi2010molecular,fazelabdolabadi2010prediction}: 
\begin{equation}
    c \: (T,P)  = \sqrt{\frac{v C_\mathit{P} (T,P)}{M C_\mathit{V} (T,V) \beta_\mathit{T} (T,P) }}
    \label{eqn:ss}
\end{equation}
\noindent where $v$ is the molar volume of the pure component or the mixture, and $M$ is the molar mass of the pure component or the mixture. For a mixture, $M$ can be calculated from the pure component molar mass:
\begin{equation}
M=\sum_{i}^{n} x_i M_i
\end{equation}
where $n$ is the number of components present in the mixture, $x_i$ and $M_i$ are the mole fraction and molar mass of each component present in the mixture. To calculate $C_\mathit{V}$, $C_\mathit{P}$, and $\beta_\mathit{T}$, the derivatives of internal energy, volume, and enthalpy with respect to temperature and pressure have to be determined \cite{lagache2001prediction, rahbari2021effect}.
\begin{align}
C_\mathit{V} (T, V) & = \left(\frac{\partial \langle U \rangle}{\partial T}\right)_\mathit{V}  
\label{eqn:CV} \\[10pt]
C_\mathit{P} (T,P) & = \left(\frac{\partial\langle H \rangle}{\partial T} \right)_P 
\label{eqn:CP} \\[10pt]
\beta_{T} (T,P) & = - \frac{1}{\langle V \rangle}\left( \frac{\partial \langle V \rangle}{\partial P}\right)_{T} 
\label{eqn:beta} 
% = \frac{1}{\langle \rho \rangle}\left( \frac{\partial \langle \rho \rangle}{\partial P}\right)_{T} 
\end{align}
\noindent where $U$ and $H$ are the internal energy and enthalpy of the system, respectively, and $\langle ... \rangle$ denotes the ensemble average of an ensemble. The internal energy ($U$ = $U^{\mathrm{internal}}$ + $U^{\mathrm{external}}$) and enthalpy ($H$ = $U^{\mathrm{internal}}$ + $U^{\mathrm{external}}$ + $K$ + $PV$) in \cref{eqn:CV} and \cref{eqn:CP} includes the kinetic energy term ($K$) in addition to the potential energy contribution from intramolecular molecular interaction (interaction inside molecules which is denoted as $U^{\mathrm{internal}}$) and intermolecular interactions (interaction between molecules which is denoted as $U^{\mathrm{external}}$). Hence, the $C_\mathit{V}$ and $C_\mathit{P}$ have been split into ideal and residual contributions. Following the work of Lagache et al. \cite{lagache2001prediction} we can write,   
\begin{equation}
    C_\mathit{V} (T, V) = C^{\mathrm{ideal}}_\mathit{V} (T) + C^{\mathrm{residual}}_\mathit{V} (T,V) =  \left(\frac{\partial \langle U^{\mathrm{ideal}} \rangle}{\partial T}\right)_\mathit{V} +  \left(\frac{\partial \langle U^{\mathrm{residual}} \rangle}{\partial T}\right)_\mathit{V}
\end{equation}
\begin{equation}
    C_\mathit{P} (T,P) = C^{\mathrm{ideal}}_\mathit{P} (T) + C^{\mathrm{residual}}_\mathit{P} (T,P) = \left(\frac{\partial \langle H^{\mathrm{ideal}} \rangle}{\partial T}\right)_\mathit{P} +  \left(\frac{\partial \langle H^{\mathrm{residual}} \rangle}{\partial T}\right)_\mathit{P}
\end{equation}
\noindent Heat capacities $C^{\mathrm{ideal}}_V$ = $\left(\frac{\partial \langle U^{\mathrm{ideal}} \rangle}{\partial T}\right)_\mathit{V}$ and $C^{\mathrm{ideal}}_P$ = $\left(\frac{\partial \langle H^{\mathrm{ideal}} \rangle}{\partial T}\right)_\mathit{P}$ can be obtained from the standard thermodynamic databases \cite{moran,poling2004properties} or from quantum mechanical calculations. In this study, the latter approach was used to calculate $C^{\mathrm{ideal}}_\mathit{V} (T)$ and $C^{\mathrm{ideal}}_\mathit{P} (T)$ using the Gaussian09 software \cite{frisch2009gaussian09} with the B3LYP theory and a 6-31G(d,p) basis set. The derivatives $\frac{\partial \langle U^{\mathrm{residual}} \rangle}{\partial T}$ and $\frac{\partial \langle H^{\mathrm{residual}} \rangle}{\partial T}$ required to calculate $C_\mathit{V}$ and $C_\mathit{P}$ are computed from fluctuations in \textit{NVT} and \textit{NPT} ensemble, respectively \cite{lagache2001prediction, fazelabdolabadi2010molecular,fazelabdolabadi2010prediction,rahbari2021effect}.
\begin{align}
C^{\mathrm{residual}}_\mathit{V} (T,V)  & =  \frac{1}{k_{\mathrm{B}}T^2} \left( \langle U^{\mathrm{external}} \hat{U} \rangle   
- \langle U^{\mathrm{external}} \rangle \langle \hat{U} \rangle \right)
\label{cv_res} \\[10pt]
C_P^{{\mathrm{residual} }}(T, P) & = \frac{1}{k_{\mathrm{B}} T^2}\left[\left\langle U^{{\mathrm{external} }} \hat{H}\right\rangle-\left\langle U^{{\mathrm{external} }}\right\rangle\langle\hat{H}\rangle  +P(\langle V \hat{H}\rangle-\langle V\rangle\langle\hat{H}\rangle)\right]-N k_{\mathrm{B}} \label{eqn:cp_res}
\end{align}
\noindent where $k_{\mathrm{B}}$ is the Boltzmann constant, $N$ is the number of molecules in the system, $\hat{U} $ ($\hat{U} $ = $U^{\mathrm{internal}}$ + $U^{\mathrm{external}}$) is the configurational energy, and $\hat{H}$ ($\hat{H}$ = $U^{\mathrm{internal}}$ + $U^{\mathrm{internal}}$ + $PV$) is the configurational enthalpy of the system. The molar heat capacities ($c_\mathit{P}$ and $c_\mathit{V}$) can be obtained from,
\begin{equation}
\frac{c_\mathit{P}}{C_\mathit{P}} = \frac{c_\mathit{V}}{C_\mathit{V}} = \frac{N_\text{A}}{N}
\label{conv}
\end{equation}
\noindent where $N_\text{A}$ is Avogadro's number and $N$ is the number of molecules in the system. The derivative $\left( \frac{\partial \langle V \rangle}{\partial P}\right)_{T}$ is required to calculate $\beta_{T}$ defined in \cref{eqn:beta}, and is computed using the following fluctuation formula \cite{allen2017computer},
\begin{equation}
     \left( \frac{\partial \langle V \rangle}{\partial P}\right)_{T} = \frac{1}{k_{\mathrm{B}} T} \left[ \langle V \rangle^2 - \langle  V^2 \rangle \right]
    \label{eqn:<v>}
\end{equation}
Analogous to the computation of the speed of sound, the calculation of the Joule-Thompson coefficient requires heat capacity at constant pressure ($c_\mathit{P}$) and thermal expansion coefficient ($\alpha_P$).
\begin{equation}
     \mu_{\mathrm{JT}} = \frac{V}{C_P} \left[ T\alpha_P - 1 \right]
      \label{eqn:JT}
\end{equation}

\noindent The value of $\alpha_P$ is computed from the derivative of volume with respect to temperature, 
\begin{equation}
\alpha_{P} (T,P) = \frac{1}{\langle V \rangle}\left( \frac{\partial \langle V \rangle}{\partial T}\right)_{P}
\label{eqn:alpha}
% =  - \frac{1}{\langle \rho \rangle} \left( \frac{\partial \langle \rho \rangle}{\partial T}\right)_{P}
\end{equation}
The derivative $\left( \frac{\partial \langle V \rangle}{\partial T}\right)_{P}$ which is required to calculate $\alpha_{T}$ is computed using \cite{lagache2001prediction,rahbari2021effect,aimoli2014force},
\begin{equation}
    \frac{\partial \langle V \rangle}{\partial T} =\frac{1}{k_\mathrm{B}T^2} \left[\langle V\hat{H} \rangle - \langle \hat{H} \rangle \langle V \rangle\right]
\end{equation}
\noindent It is important to note that $C_\mathit{V}$ and $\alpha_{P}$ can be computed indirectly using the thermodynamic relations, \cite{callen1998thermodynamics}
\begin{align}
C_V & = C_P-\frac{T\langle V\rangle \alpha_P^2}{\beta_T}
\label{CV-R} \\[10pt]
\alpha_{\mathit{P}} & =\sqrt{\left(\left(C_{\mathit{P}}-C_{\mathit{V}}\right) \frac{\beta_{\mathit{T}}}{T\langle V\rangle}\right)}
\end{align}
\noindent The proofs of the mathematical derivation of equations to compute $C_P$, $C_V$, $\beta_T$, $c$, $\alpha_P$, and $\mu_{\mathrm{JT}}$ from simulations are provided in Sections S3-S8 of the Supporting Information.

\section{Simulation details}
\label{section-3}
All force field-based MC simulations were performed using the open-source software Brick, which uses the Continuous Fractional Component Monte Carlo (CFCMC) method \cite{hens2020brick,polat2021new,rahbari2021recent,shi2007continuous,shi2008improvement} to calculate thermodynamics properties. Force field-based MD simulations were performed in the Large-scale Atomic/Molecular Massively Parallel Simulator (LAMMPS: version August 2023) package \cite{thompson2022lammps} to compute viscosities. All molecules were described with site-based conventional intermolecular potentials with point charges centered at the atom or at a dummy site. The pairwise-additive 12-6 Lennard-Jones (LJ) interaction potentials are used to model interactions:
\begin{equation}
u\left(r_{i j}\right)=4 \epsilon_{i j}\left[\left(\frac{\sigma_{i j}}{r_{i j}}\right)^{12}-\left(\frac{\sigma_{i j}}{r_{i j}}\right)^6\right]+\frac{q_i q_j}{4 \pi \epsilon_0 r_{i j}}\label{12-6}
\end{equation}
The Lorentz-Berthelot mixing rules were used to compute LJ interactions for dissimilar atoms \cite{frenkel2023understanding,allen2017computer}:
\begin{equation}
    \sigma_{i j}=\frac{1}{2}\left(\sigma_{i i}+\sigma_{j j}\right)\label{sigma-LJ}
\end{equation}
\begin{equation}
    \varepsilon_{i j}=\left(\varepsilon_{i i} \varepsilon_{j j}\right)^{1 / 2} \label{epsilon-LJ}
\end{equation}
\noindent \textcolor{black}{Different force fields for \ce{CO2} are available in the literature. These range from simple single-site force fields like the Higashi model \cite{higashi1998diffusion} and Statistical Associating Fluid Theory (SAFT)-$\gamma$ \cite{avendano2011saft} to more complex three-site force fields such as the Transferable Potentials for Phase Equilibria (TraPPE)-rigid \cite{potoff2001vapor}, TraPPE-flex \cite{perez2010molecular}, Elementary Physical Model 2 (EPM2) \cite{harris1995carbon}, Zhang model \cite{zhang2005optimized}, and Cygan model \cite{cygan2012molecular}. Aimoli et al. \cite{aimoli2014force} investigated the performance of seven \ce{CO2} force fields (TraPPE-rigid, TraPPE-flex, EPM2, Zhang model, Cygan model, Higashi model, and SAFT-$\gamma$) on density and second derivative thermodynamic properties of \ce{CO2} up to \SI{900}{\kelvin} and 1000 bar. Alimoli et al. \cite{aimoli2014force} found that TrappE-rigid, EPM2, the Zhang model, and SAFT-$\gamma$ produced nearly identical densities of \ce{CO2} but the SAFT-$\gamma$ force field predicted second derivative properties less accurately than TrappE-rigid, EPM2, and Zhang model compared to reference data from National Institute of Standards and Technology (NIST). Aimoli et al. \cite{aimoli2014force} also investigated the TraPPE and SAFT-$\gamma$ single-site force fields for \ce{CH4} and found that the TraPPE force field performed the best compared to NIST reference data. Similar to \ce{CO2}, several force fields for \ce{N2} can be found in the literature, such as TraPPE \cite{potoff2001vapor}, K\"{o}ster et al. model \cite{koster2018molecular}, Murthy et al. model \cite{murthy1983electrostatic}, Galassai and Tildesley model \cite{galassi1994phase}. Force fields from Galassai and Tildesley \cite{galassi1994phase} and Murthy et al. \cite{murthy1983electrostatic} were not optimized for VLE calculations \cite{potoff2001vapor}. The force fields TraPPE \cite{potoff2001vapor} and K\"{o}ster et al. \cite{koster2018molecular} are three-site models with quadrupole moment but differ in their parameterization. Rahbari et al. \cite{rahbari2021effect} compared the performance of five different force fields for \ce{H2} (Cracknell \cite{cracknell2001molecular}, Buch \cite{buch1994path}, Hirschfelder et al. \cite{hirschfelder1964molecular}, Marx and Nielaba \cite{marx1992path}, and K\"{o}ster et al. \cite{koster2018molecular}). These authors found that the force field developed by K\"{o}ster et al. \cite{koster2018molecular} best predicted the second derivative thermodynamic properties with the least deviation compared to data obtained from REFPROP \cite{lemmon2018nist}. The TraPPE \cite{potoff2001vapor,martin1998transferable} force field was used in this study for \ce{CO2}, \ce{N2}, and \ce{CH4} molecules since the primary objective in developing TraPPE force field is to predict the thermophysical properties for a wide range of state conditions and compositions. Hydrogen is simulated using the K\"{o}ster et al. \cite{koster2018molecular} model. The force field from Garc\'{i}a-P\'{e}rez et al. \cite{garcia2008unraveling} commensurate with tail corrections was used for monatomic non-polar argon molecule.} Molecular models of \ce{H2}, \ce{Ar}, and \ce{CH4} are single-site models consisting of a single Lennard-Jones (LJ) interaction site with no point charges, whereas \ce{CO2} and \ce{N2} are three-site models (including the dummy charge site for \ce{N2}) with point charges. The force field parameters of \ce{CO2}, \ce{Ar}, \ce{N2}, \ce{H2}, and \ce{CH4} are listed in the Section S1 of the Supporting Information. All molecules were treated as rigid objects. LJ interactions in a simulation box have a cutoff radius of \SI{12}{\angstrom} with analytic tail corrections \cite{frenkel2023understanding}. Periodic boundary conditions were imposed in all directions. The Ewald summation is used to compute the electrostatic energy due to point charges. To minimize computation expense, the cutoff radius of real space electrostatic interactions is chosen to limit the \textit{k}-vectors (to a maximum of \textit{k} $=$ $8$) in Fourier space, with an accuracy of $10^{-6}$. For instance, we chose a cutoff radius in real space as \SI{12}{\angstrom} with a damping parameter of $\alpha$ $=$ $0.2650$ $\text{\AA}^{-1}$ for a box of size \SI{30}{\angstrom}, whereas for a box of size \SI{40}{\angstrom} we chose cutoff radius in real space as \SI{16}{\angstrom} with a damping parameter of $\alpha$ $=$ $0.1960$ $\text{\AA}^{-1}$. For each condition (concentration, temperature, and pressure), 10 independent simulations are performed, and each simulation starts with a different initial configuration. These 10 simulations are divided into 5 blocks from which average values and uncertainties of thermodynamic and transport properties are calculated. The mean and standard deviation of 5 blocks is the average value and uncertainty of a thermodynamic or transport property. The aforementioned force fields and simulation details used for MC and MD simulations are exactly the same.

MC simulation of Gibbs Ensemble (GE) in \textit{NVT} and \textit{NPT} version is the most convenient way to perform phase equilibria calculations \cite{panagiotopoulos1994molecular,panagiotopoulos2000monte,dinpajooh2015accurate}. In GE, two simulation boxes are considered: one represents the liquid phase, and the other represents the gas phase. The simulation boxes are allowed to exchange energy, volume, and molecules. In a dense liquid simulation box, the insertion of a molecule in a single step is impeded due to the low probability of finding a cavity to accommodate a molecule, and the deletion of a molecule in a single step leaves the simulation box with a high energy penalty to form a new configuration \cite{rahbari2021recent}. The CFCMC method \cite{shi2007continuous,shi2008improvement,hens2020brick,polat2021new,rahbari2021recent} overcomes this drawback by gradual insertion and removal of so-called fractional molecules by which the surrounding whole molecules can adapt simultaneously by performing trial moves related to fractional molecules besides thermalization trial moves such as translations, rotations, and volume changes. In CFCGE, two simulation boxes with indistinguishable whole molecules and fractional molecules (fractional molecules can be in either of the simulation boxes but one per component type) are used for simulating the phase coexistence. The interaction of the distinguishable fractional molecule of a component type $i$ with a whole molecule is scaled with a coupling parameter $\lambda_i$ $\in$ $[0,1]$ \cite{poursaeidesfahani2016direct}. The trial moves related to fractional molecules are randomly changing the value of $\lambda$ while keeping the orientation and position of all the molecules constant, insertion of a fractional molecule in another simulation box at a randomly selected orientation and position while keeping the orientation and position of the whole molecules constant, and changing the identity of a fractional molecule in a simulation box to a whole molecule while simultaneously transforming the randomly selected whole molecule to a fraction molecule in another simulation box \cite{poursaeidesfahani2017computation}. Further details specific to phase equilibria calculation in the CFCGE can be found elsewhere \cite{rahbari2021recent,poursaeidesfahani2016direct}. The VLE of pure components (unary) systems \ce{CO2}, \ce{Ar}, \ce{N2}, \ce{H2}, and \ce{CH4}, were computed in the $\mathit{NVT}$ version of the CFCGE. The phase equilibria (\textit{Pxy} diagram) of binary mixtures which include \ce{CO2}/\ce{Ar}, \ce{CO2}/\ce{N2}, \ce{CO2}/\ce{H2}, and \ce{CO2}/\ce{CH4} are computed in the $\mathit{NPT}$ version of the CFCGE. The simulation box sizes and initial distribution of whole molecules between the simulation boxes were specified based on the experimental data of the state point for both the unary and binary systems. For example, phase equilibria computation of \ce{CO2} and \ce{Ar} binary system at 105 bar, we chose 360 \ce{CO2} molecules and 140 \ce{Ar} molecules for the intended liquid box of size \SI{30}{\angstrom} and 180 \ce{CO2} molecules, and 320 \ce{Ar} molecules for the intended gas box of size \SI{40}{\angstrom}. For a pressure of 35 bar, we chose 470 \ce{CO2} molecules and 30 \ce{Ar} molecules for the intended liquid box of size \SI{30}{\angstrom} and 325 \ce{CO2} molecule and 175 \ce{Ar} molecules for the intended gas box of size \SI{40}{\angstrom}. For simulating the phase coexistence of unary and binary systems, an equilibration run of $5$ x $10^{4}$ and $1$ x $10^{5}$ MC cycles was performed, respectively. The number of trial moves in an MC cycle in Brick-CFCMC equals the total number of molecules in the simulation box with a minimum of 20. Following the equilibration run, a production run of $1$ x $10^{5}$ cycles was performed for unary systems and $2$ x $10^{5}$ cycles for binary systems to compute the coexistence densities and mole fractions of the components, respectively.

To calculate thermodynamic properties: density ($\rho$), isothermal compressibility ($\beta_\mathit{T}$), thermal expansion coefficient ($\alpha_P$), molar heat capacity at constant volume ($c_V$), molar heat capacity at constant pressure ($c_P$), Joule-Thompson coefficient ($\mu_{\mathrm{JT}}$), and speed of sound ($c$) MC simulations were performed without fractional molecules. The values of $c_P$, $\beta_T$, $\mu_{\mathrm{JT}}$, and $\alpha_P$ were computed using the fluctuation equations in the $\mathit{NPT}$ ensemble defined in \cref{eqn:beta,eqn:alpha,eqn:CP,eqn:JT}, respectively. To compute the $c_V$ and consequently the speed of sound ($c$), it is important to extract the ensemble-averaged volume ($\langle V \rangle$) from the $\mathit{NPT}$ ensemble that reflects the same state for performing simulation in the $\mathit{NVT}$ ensemble with the same number of molecules. The simulations were performed with 300 molecules, irrespective of the state and system. An equilibration of $5$ x $10^{4}$ MC cycles is performed to equilibrate the system successively, and $10^{6}$ production runs are performed for each simulation.

In MD simulations, initial configurations of molecules in the cubic simulation boxes are constructed using PACKMOL \cite{martinez2009packmol} and fftool \cite{agilio_padua_2021_4701065}. Periodic boundary conditions are applied to simulation boxes in all directions. A cutoff radius of \SI{12}{\angstrom} is used for LJ interactions with analytic tail corrections. All molecules are treated as rigid bodies, and Newton's equations of motion are integrated using the velocity-Verlet algorithm with a time step of 0.5 fs. To thermostat and barostat the system, the No\'se-Hoover type is used with coupling constants of 0.1 ps and 1 ps, respectively. The Particle-Particle Particle-Mesh (PPPM) method is used to handle long-range electrostatic interactions with a cutoff radius of \SI{12}{\angstrom} and $10^{-6}$ accuracy. Shear viscosities ($\eta$) are calculated by performing an Equilibrium Molecular Dynamics (EMD) simulation by using the On-the-fly Computation of Transport Properties (OCTP) plugin in LAMMPS. The OCTP combines the Einstein relations with an order-$n$ algorithm to calculate viscosity. Additional details about the OCTP's computation of transport properties can be found elsewhere \cite{jamali2019octp}. System sizes chosen to evaluate $\eta$ of pure and multicomponent mixtures were 100 molecules for very diluted gas phase, 300 molecules for the gas phase (40 to 80 bar), and 400 molecules for the liquid and supercritical phases ($\geq$ 100 bar). The various steps involved in computing $\eta$ for one independent simulation are as follows: first, a simulation is carried out in the $\mathit{NPT}$ ensemble (0.5 ns equilibration run and 1 ns production run) to compute the ensemble average volume ($\langle V \rangle$). Next, the simulation box is scaled according to the computed value of $\langle V \rangle$, and this system is then used to perform simulation in the $\mathit{NVT}$ ensemble (0.5 ns equilibration run and 1 ns production run) to compute the average total energy of the system ($\langle E \rangle$). Finally, the ensemble average total energy is used to scale to the kinetic energy of the system to perform simulation in $\mathit{NVE}$ ensemble. Viscosities are calculated in the \textit{NVE} ensemble, ensuring that the thermostat and barostat have no effect on the results. In $\mathit{NVE}$ ensemble, a production run of 5 ns is simulated to compute $\eta$.

\section{Results and discussion}
\label{section-4}
This section shows and discusses the thermodynamic and transport properties of single and multicomponent \ce{CO2} systems. All properties of interest computed in this work from molecular simulations are compared with data sets generated from the NIST REFPROP database version-10.0 \cite{lemmon2018nist} except when stated otherwise. EoS and correlation models used by REFPROP for computing pure components (\ce{CO2}, \ce{N2}, \ce{Ar}, \ce{H2}, and \ce{CH4}) thermodynamic and transport properties are listed in Table S11 of the Supporting Information. For multicomponent systems, the Groupe Europ\`een de Recherches Gazi\`eres (GERG-2008) EoS \cite{kunz2012gerg} is chosen in REFPROP. The GERG-2008 EoS is less accurate for unary systems \cite{lemmon2018nist}. Therefore, it is only used to compute the thermodynamic properties of multicomponent systems. \textcolor{black}{The GERG-2008 EoS was originally developed for natural gas mixtures containing 21 components, including \ce{CO2}, \ce{N2}, \ce{Ar}, \ce{H2}, and \ce{CH4}. Multicomponent mixtures containing high \ce{CO2} concentrations and low levels of impurities were not the main focus for the development of the GERG-2008 EoS \cite{kunz2012gerg,gernert2016eos}. The quality and quantity of experimental data used in developing the GERG-2008 EoS limits its accuracy \cite{kunz2012gerg}. Thermophysical properties computed from molecular simulations will help in optimizing the EoS \cite{ramdin2016computing}.} \textcolor{black}{Viscosities computed from MD simulations of multicomponent systems are compared with those obtained from the Extending Corresponding States (ECS) model available in REFPROP \cite{huber2022nist}. Further details specific to the ECS model can be found elsewhere \cite{huber2022nist}.} Since the experimental data for multicomponent systems are scarce, and simulating an exact composition as experiments is impossible with a small system of only 300 molecules, we opted to compare results from molecular simulations with those obtained from EoS. The numerical data used to generate all plots is provided in Sections S3, S4, and S14-S17 of the Supporting Information. The deviations of properties computed from simulations with respect to REFPROP data sets are computed using:

\begin{equation}
    \text{Relative Deviation \%} = \left| \frac{\chi^{\mathrm{REFPROP}} - \chi^{\mathrm{Simulation}}}{\chi^{\mathrm{REFPROP}}} \right| \times 100 \%
    \label{eqn:RD}
 \end{equation}
 
\noindent where $\chi^{\mathrm{Simulation}}$ and $\chi^{\mathrm{REFPROP}}$ are the property of interest computed from simulations and REFPROP, respectively. For the sake of clarity, plots for temperatures \SI{253}{\kelvin} and \SI{313}{\kelvin} are shown for binary systems, and plots for a temperature \SI{253}{\kelvin} are shown for multicomponent systems despite properties of interest were estimated for four different temperatures: \SI{253}{\kelvin}, \SI{273}{\kelvin}, \SI{293}{\kelvin}, and \SI{313}{\kelvin}.

\subsection{Thermodynamic properties}

\subsubsection{Phase equilibria}
The VLE of pure components (\ce{CO2}, \ce{Ar}, \ce{N2}, \ce{H2}, and \ce{CH4}) computed from the CFCMC simulations in the \textit{NVT} version are compared to REFPROP database. The computed VLE curves compared to REFPROP database are shown in Section S2 of the Supporting Information. Critical temperatures ($T_c$) and densities ($\rho_c$) of all pure components computed from simulations using the law of rectilinear diameters and EoS models are listed in \cref{us}. Our results show that the computed liquid and vapor densities are in excellent agreement with respect to REFPROP dataset, except for the liquid densities of hydrogen. Deviations of the computed liquid densities and, in turn, the $T_c$ and $\rho_c$ of hydrogen with respect to REFPROP are due to the domination of quantum effects at low temperatures \cite{koster2018molecular}. This work focuses on temperatures significantly higher ($>$ \SI{250}{\kelvin}) than \ce{H2} VLE temperatures, hence, only the gas densities were considered for the validation of \ce{H2} force field. 

\begin{figure}[tbh!]
\centering
\begin{subfigure}{0.5\textwidth}
\centering
\caption{}
\includegraphics[width=\textwidth]{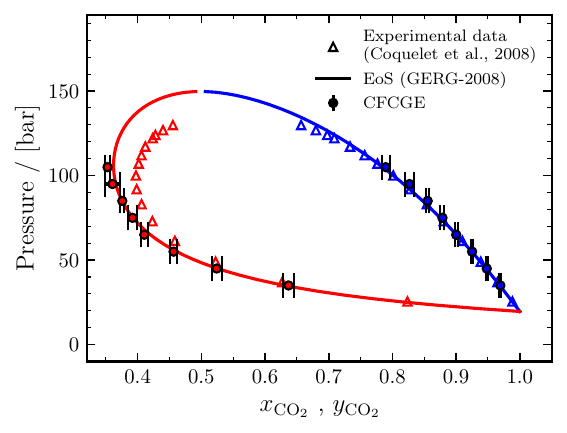}
\end{subfigure}%
\begin{subfigure}{0.5\textwidth}
\centering
\caption{}
\includegraphics[width=\textwidth]{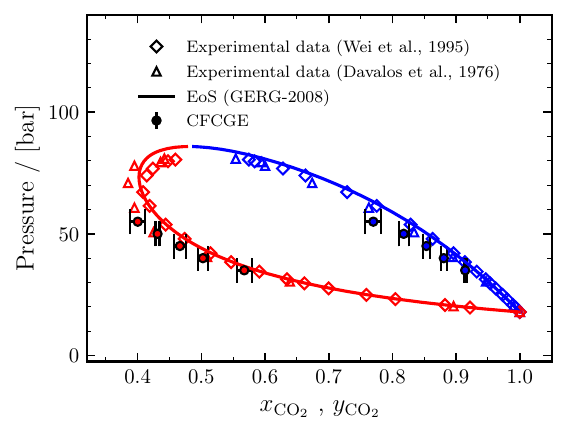}
\end{subfigure}
\begin{subfigure}{0.5\textwidth}
\centering
\caption{}
\includegraphics[width=\textwidth]{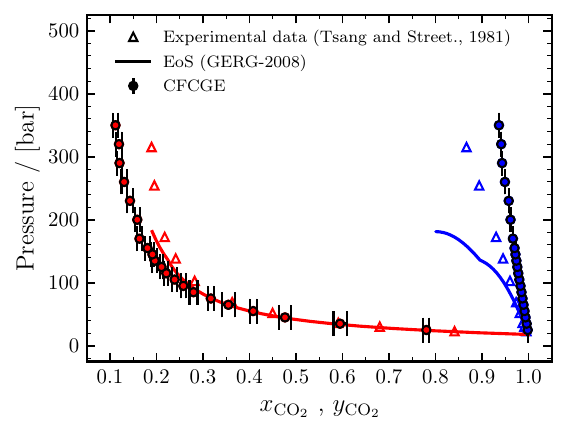}
\end{subfigure}%
\begin{subfigure}{0.5\textwidth}
\centering
\caption{}
\includegraphics[width=\textwidth]{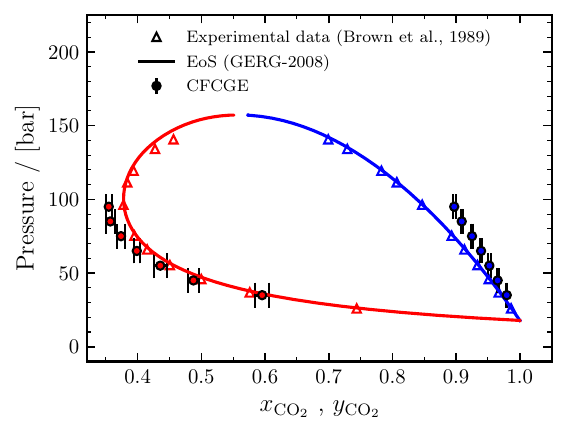}
\end{subfigure}
\caption{Comparison of binary VLE (\textit{Pxy} diagram), i.e., the bubble points (blue symbols and lines) and dew points (red symbols and lines) of (a) \ce{CO2}/\ce{Ar}, (b) \ce{CO2}/\ce{CH4},(c) \ce{CO2}/\ce{H2}, and (d) \ce{CO2}/\ce{N2} mixtures computed from CFCGE simulations with experimental data \cite{coquelet2008isothermal,davalos1976liquid,wei1995vapor+,tsang1981phase,brown1989vapor} and the GERG-2008 EoS \cite{kunz2012gerg}. The simulations are performed at \SI{253.28}{\kelvin} for \ce{CO2}/\ce{Ar} mixtures and at \SI{250}{\kelvin} for \ce{CO2}/\ce{CH4}, \ce{CO2}/\ce{H2}, and \ce{CO2}/\ce{N2} mixtures.}
\label{fig:BM_VLE}
\end{figure}

\cref{fig:BM_VLE} shows the phase equilibria (\textit{Pxy} diagram) of four binary mixtures comprising of \ce{CO2}, \ce{Ar}, \ce{N2}, \ce{H2}, and \ce{CH4} computed from CFCGE simulations (in the \textit{NPT} version) compared with experimental data and the GERG-2008 EoS. The \textit{Pxy} diagram, i.e., the bubble and dew points of \ce{CO2}/\ce{Ar} mixtures, is computed at \SI{253.28}{\kelvin} and compared with experimental data of Coquelet et al. \cite{coquelet2008isothermal} and for \ce{CO2}/\ce{CH4} mixtures, simulations were performed at \SI{250}{\kelvin} and compared to experimental data of Wei et al. \cite{wei1995vapor+} and Davalos et al. \cite{davalos1976liquid}. Similarly, \textit{Pxy} diagram of \ce{CO2}/\ce{H2} and \ce{CO2}/\ce{N2} mixtures are computed at \SI{250}{\kelvin} and validated with experimental data of Tsang and Street \cite{tsang1981phase} and Brown et al. \cite{brown1989vapor}, respectively. \cref{fig:BM_VLE} (a) shows \ce{CO2} liquid and gas mole fractions of \ce{CO2}/\ce{Ar} mixtures, computed from the CFCGE simulations compared to experimental data of Coquelet et al. \cite{coquelet2008isothermal} and the GERG-2008 EoS. Our results match well with the GERG-2008 EoS compared to the experimental data of Coquelet et al. \cite{coquelet2008isothermal}. Bubble and dew points of \ce{CO2}/\ce{CH4} mixtures (shown in \cref{fig:BM_VLE} (b)) computed from simulations agrees well with experimental data of Davalos et al. \cite{davalos1976liquid} when compared to experimental data of Wei et al. \cite{wei1995vapor+} and the GERG-2008 EoS at high pressures. The phase equilibria of \ce{CO2}/\ce{H2} binary system is shown in \cref{fig:BM_VLE} (c). REFPROP fails to converge for pressures larger than \SI{170}{\bar} when using the GERG-2008 EoS. This discrepancy seen in the GERG-2008 EoS was also reported in the work of Shin et al. \cite{shin2018evaluation}. \textcolor{black}{The bubble points obtained from the GERG-2008 EoS using REFPROP provide poor estimates when compared to the experimental data of Tsang and Street \cite{tsang1981phase}.} Mole fractions computed from simulations agree fairly with experimental data of Tsang and Street \cite{tsang1981phase}, and less than 5\% relative deviation was noticed at high pressures. Dew points of \ce{CO2}/\ce{N2} mixtures shown in \cref{fig:BM_VLE} (d) have a reasonable agreement with EoS and experimental data of Brown et al. \cite{brown1989vapor}, but the computed bubble points have a maximum relative deviation of 3\% at high pressures. \textcolor{black}{At high pressures, the dew and bubble points computed from CFCGE simulations agree moderately with the GERG-2008 EoS and experiments for all systems shown in \cref{fig:BM_VLE}. This is because the mixing rules used in simulations did not include binary interaction parameters.} Computing bubble and dew points close to the critical pressure is challenging, even with a system size of 1000 molecules (the system size chosen in this study to compute the phase equilibria). In principle, one can perform simulations with large system sizes for pressures in the neighborhood of the critical pressure at the cost of much larger computations. Nevertheless, it is impossible to compute accurate bubble and dew points very close to critical pressure since simulation boxes may switch identities, which complicates ensemble averaging. Similarly, the GERG-2008 EoS model fails to predict reliable results for pressures close to the critical pressure due to the unavailability of experimental data to develop the GERG-2008 EoS \cite{gernert2016eos}.

\subsubsection{Densities \texorpdfstring{($\rho$)}{Lg}}

\begin{figure}[H]
\centering
\begin{subfigure}{0.49\textwidth}
\caption{}
\includegraphics[width=\textwidth]{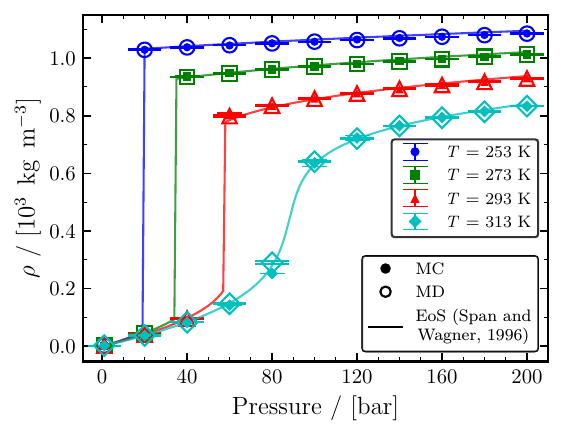}
\end{subfigure}
\begin{subfigure}{0.49\textwidth}
\caption{}
\includegraphics[width=\textwidth]{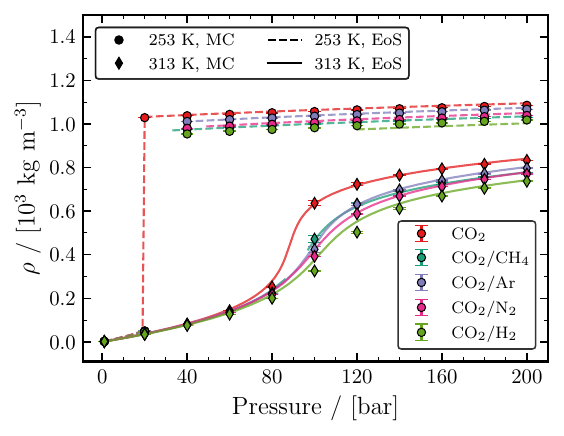}
\end{subfigure}
\begin{subfigure}{0.7\textwidth}
\caption{}
\includegraphics[width=\textwidth]{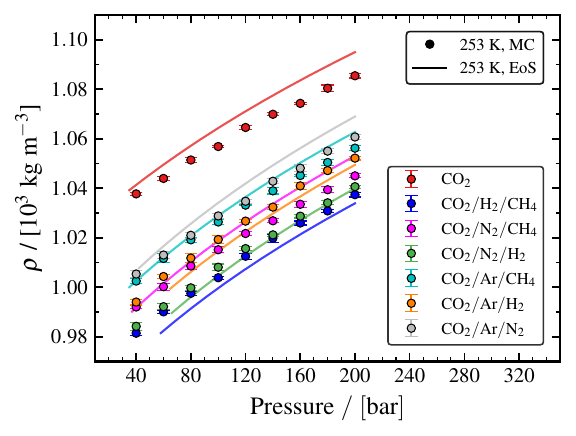}
\end{subfigure}
\caption{Computed densities as a function of temperature and pressure. (a) shows the calculated densities of pure \ce{CO2} from MC simulations (closed symbols), MD simulations (open symbols), and the Span and Wagner EoS  \cite{span1996new} (solid lines) for temperatures: \SI{253}{\kelvin}, \SI{273}{\kelvin}, \SI{293}{\kelvin}, and \SI{313}{\kelvin}. (b) shows densities of binary mixtures with 95 mol\% of \ce{CO2} and 5 mol\% of impurities (\ce{CH4}, \ce{Ar}, \ce{N2}, and \ce{H2}) computed from MC simulations (closed symbols) and the GERG-2008 EoS \cite{kunz2012gerg} (lines) compared with densities of pure \ce{CO2} computed from MC simulations and the Span and Wagner EoS \cite{span1996new} for temperatures: \SI{253}{\kelvin} and \SI{313}{\kelvin}. (c) shows densities of ternary mixtures with 96 mol\% of \ce{CO2} and 2 mol\% impurities for each of two components (\ce{CH4}, \ce{Ar}, \ce{N2}, and \ce{H2}) computed from MC simulations (closed symbols) and the GERG-2008 EoS \cite{kunz2012gerg} (lines) compared with densities of pure \ce{CO2} computed from MC simulations and the Span and Wagner EoS \cite{span1996new} at \SI{253}{\kelvin}.}
\label{fig:density}
\end{figure}

Densities of pure \ce{CO2} computed from MC and MD simulations for temperatures \SI{253}{\kelvin}, \SI{273}{\kelvin}, \SI{293}{\kelvin}, and \SI{313}{\kelvin} and pressures ranging from 20 bar to 200 bar are shown in \cref{fig:density} (a). MC and MD simulations have an excellent agreement with the Span and Wagner EoS for all temperatures. The computed densities from MC simulations have a maximum relative deviation of ca. 0.93\% at \SI{313}{\kelvin} and 200 bar, and MD simulations have a maximum relative deviation of ca. 0.89\% at \SI{293}{\kelvin} and 200 bar, excluding conditions close to the critical point. As expected, the density decreases with temperature to a large extent after the transition from gas to liquid or supercritical fluid. \cref{fig:density} (b) shows densities of different binary mixtures computed from MC simulations compared with densities obtained from the GERG-2008 EoS for 95 mol\% of \ce{CO2} and 5 mol\% of one of impurities (\ce{CH4}, \ce{Ar}, \ce{N2}, and \ce{H2}) for temperatures \SI{253}{\kelvin} and \SI{313}{\kelvin}. The computed densities of \ce{CO2}/\ce{N2}, \ce{CO2}/\ce{CH4}, \ce{CO2}/\ce{Ar} and \ce{CO2}/\ce{H2} binary mixtures from MC and MD are in close agreement with densities obtained from the GERG-2008 EoS and has a maximum relative deviation of ca. 4.4\% (in MC simulation at \SI{313}{\kelvin} and 60 bar), ca. 4.3\% (in MD simulations at \SI{253}{\kelvin} and 80 bar), ca. 2.1\% (in MC simulations at \SI{253}{\kelvin} and 40 bar), ca. 6.6\% (in MD simulations at \SI{313}{\kelvin} and 60 bar), respectively, excluding state points close to the critical point. A comprehensive analysis of the gas phase densities of binary mixtures with 5 mol\% of one of the impurities found that impurities do not significantly change the gas phase densities. The influence of impurities on densities is consistent with the molecular weight of the mixtures. A mixture with \ce{H2} as an impurity decreases the density to a large extent, followed by \ce{Ar}, \ce{N2}, and \ce{CH4}. Densities were also computed for 1 mol\% and 10 mol\% impurities at temperatures \SI{253}{\kelvin}, \SI{273}{\kelvin}, \SI{293}{\kelvin}, and \SI{313}{\kelvin} from MC simulations and MD simulations. The computed densities are provided in Tables S73-S167 of the Supporting Information. Densities of all binary mixtures shown in Section S15 of the Supporting Information, densities decrease with increasing mol\% of impurity compared to densities of pure \ce{CO2} irrespective of the temperature. \cref{fig:density} (c) shows the liquid densities of ternary mixtures with 96 mol\% \ce{CO2} and 2 mol\% impurities for each of the two components (\ce{CH4}, \ce{Ar}, \ce{N2}, and \ce{H2}) at \SI{253}{\kelvin}. \textcolor{black}{The liquid densities of ternary mixtures may appear to have some deviation when compared to the GERG-2008 EoS due to the reduced axis range seen in \cref{fig:density} (c). The maximum relative deviation of ternary mixtures liquid densities in \cref{fig:density} (c) was ca. 0.78\% for \ce{CO2}/\ce{Ar}/\ce{N2} and \ce{CO2}/\ce{N2}/\ce{CH4} mixtures at 200 bar. The liquid densities of ternary mixtures were observed to have good agreement with the GERG-2008 EoS.} Similar to binary mixtures, ternary mixtures with the least molecular weight tend to have lesser densities when compared to other ternary mixtures. For example, a ternary mixture with 2 mol\% \ce{H2} and 2 mol\% \ce{CH4} as impurities, which have the lowest molecular weight among other ternary mixtures, have the lowest densities when compared to densities of other ternary mixtures.

\subsubsection{Thermal expansion coefficients \texorpdfstring{($\alpha_{P}$)}{Lg}}

\begin{figure}[H]
\centering
\begin{subfigure}{0.49\textwidth}
\caption{}
\includegraphics[width=\textwidth]{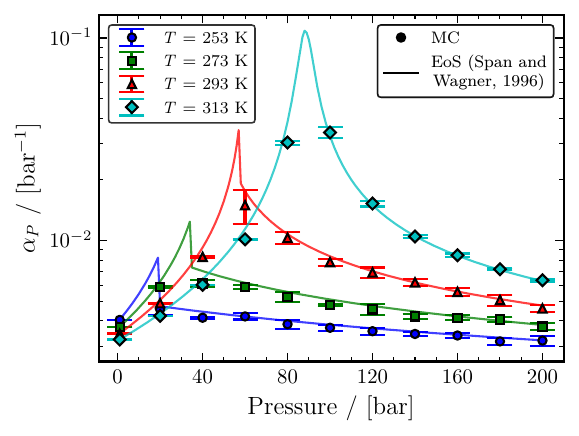}
\end{subfigure}
\begin{subfigure}{0.49\textwidth}
\caption{}
\includegraphics[width=\textwidth]{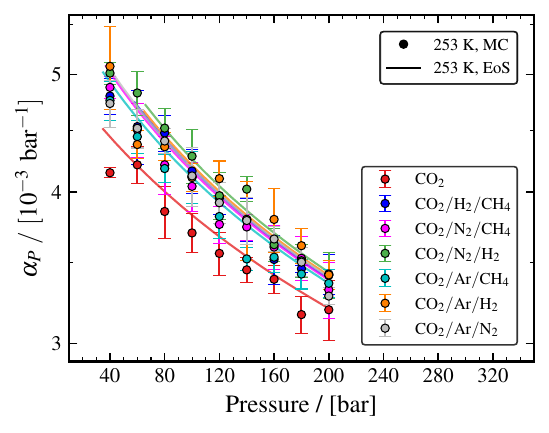}
\end{subfigure}
\begin{subfigure}{1\textwidth}
\caption{}
\includegraphics[width=\textwidth]{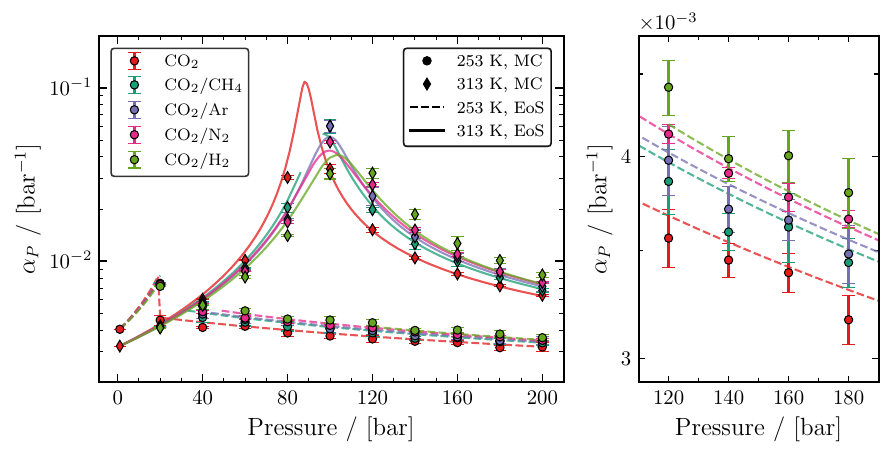}
\end{subfigure}
\caption{Computed thermal expansion coefficients as a function of temperature and pressure. (a) shows the calculated thermal expansion coefficients of pure \ce{CO2} from MC simulations (closed symbols) and the Span and Wagner EoS  \cite{span1996new} (solid lines) for temperatures: \SI{253}{\kelvin}, \SI{273}{\kelvin}, \SI{293}{\kelvin}, and \SI{313}{\kelvin}. (b) shows thermal expansion coefficients of ternary mixtures with 96 mol\% of \ce{CO2} and 2 mol\% impurities for each of two components (\ce{CH4}, \ce{Ar}, \ce{N2}, and \ce{H2}) computed from MC simulations (closed symbols) and the GERG-2008 EoS \cite{kunz2012gerg} (lines) compared with thermal expansion coefficients of pure \ce{CO2} computed from MC simulations and the Span and Wagner EoS \cite{span1996new} for temperatures: \SI{253}{\kelvin} and \SI{313}{\kelvin}. (c) shows thermal expansion coefficients of binary mixtures with 95 mol\% of \ce{CO2} and 5 mol\% of impurities (\ce{CH4}, \ce{Ar}, \ce{N2}, and \ce{H2}) computed from MC simulations (closed symbols) and the GERG-2008 EoS \cite{kunz2012gerg} (lines) compared with thermal expansion coefficients of pure \ce{CO2} computed from MC simulations and the Span and Wagner EoS \cite{span1996new} at \SI{253}{\kelvin}.}
\label{alpha}
\end{figure}

Thermal expansion coefficient ($\alpha_{P}$) computed from MC simulations using \cref{eqn:alpha} in the \textit{NPT} ensemble for pure \ce{CO2}, binary \ce{CO2} mixtures, and ternary \ce{CO2} rich mixtures with different impurities are shown as a function of temperature and pressure in \cref{alpha}. $\alpha_{P}$ of pure \ce{CO2} computed at temperatures \SI{253}{\kelvin}, \SI{273}{\kelvin}, \SI{293}{\kelvin}, and \SI{313}{\kelvin} are shown in \cref{alpha} (a). $\alpha_{P}$ of pure \ce{CO2} increases with temperature, whereas with increasing pressure, the value of $\alpha_{P}$ increases till it reaches a maximum value close to its saturation pressure and then decreases with increasing pressure. MC simulations closely predicted the peak of $\alpha_{P}$. Thermal expansion coefficients of pure \ce{CO2} computed at conditions close to its saturation pressure were found to have large uncertainties and deviations when compared to the Span and Wagner EoS. Excluding conditions close to the critical point, $\alpha_{P}$ of pure \ce{CO2} computed from MC simulations were in good agreement with the Span and Wagner EoS for all temperatures with a maximum relative deviation of ca. 6.8\% at \SI{293}{\kelvin} and 100 bar. In binary mixtures containing 95 mol\% of \ce{CO2} and 5 mol\% of one of the impurities (\ce{CH4}, \ce{Ar}, \ce{N2}, and \ce{H2}), a similar pattern of $\alpha_{P}$ with a peak value near its saturation pressure was found, as shown in \cref{alpha} (c) at temperatures \SI{253}{\kelvin} and \SI{313}{\kelvin}. The computed $\alpha_{P}$ of binary mixtures \ce{CO2}/\ce{N2}, \ce{CO2}/\ce{CH4}, \ce{CO2}/\ce{Ar} and \ce{CO2}/\ce{H2} agrees fairly with $\alpha_{P}$ obtained from the GERG-2008 EoS and has a maximum relative deviation of ca. 13.9\% at \SI{313}{\kelvin} and 120 bar, ca. 10.2\% at \SI{313}{\kelvin} and 60 bar, ca. 8.1\% at \SI{313}{\kelvin} and 60 bar, ca. 20.4\% at \SI{313}{\kelvin} and 120 bar, respectively, excluding state points close to the critical point. State points of binary mixtures close to the critical point were identified based on large uncertainties observed in simulations and large relative deviation with the GERG-2008 EoS. The uncertainties and relative deviation of \ce{CO2}/\ce{H2} at \SI{313}{\kelvin} decrease with increasing pressure. This suggests that the maximum relative deviation for \ce{CO2}/\ce{H2} mixture observed in simulations at \SI{313} {\kelvin} and 120 bar may be near its saturation/Widom line. The binary mixture containing 5 mol\% of \ce{H2} increases the value of $\alpha_{P}$ the most in liquid and supercritical phases when compared to $\alpha_{P}$ of pure \ce{CO2}, followed by binary mixtures containing 5 mol\% of \ce{N2}, \ce{Ar}, and \ce{CH4}. In the gas phase, the presence of \ce{H2} as impurity decreases the value of $\alpha_{P}$ followed by \ce{CH4}, \ce{N2}, and \ce{Ar} compared to $\alpha_{P}$ of pure \ce{CO2}. Thermal expansion coefficients were also computed for 1 mol\% and 10 mol\% impurities at temperatures \SI{253}{\kelvin}, \SI{273}{\kelvin}, \SI{293}{\kelvin}, and \SI{313}{\kelvin} from MC simulations. The computed thermal expansion coefficients provided in Tables S73-S167 of the Supporting Information show that the value of $\alpha_{P}$ are affected based on impurities and concentration level of impurities. Thermal expansion coefficients of ternary mixtures in the liquid phase with 96 mol\% \ce{CO2} and 2 mol\% for each of the two impurities (\ce{CH4}, \ce{Ar}, \ce{N2}, and \ce{H2}) are shown in \cref{alpha} (b). The effect of a particular mixture on $\alpha_{P}$ was not discernible since the thermal expansion coefficients of all the ternary mixtures computed at \SI{253}{\kelvin} were within the limits of computed uncertainty of other ternary mixtures. Similarly, no significant differences were found between the thermal expansion coefficients of ternary mixtures obtained from the GERG-2008 EoS.

\subsubsection{Isothermal compressibilities \texorpdfstring{($\beta_{T}$)}{Lg}}

\begin{figure}[H]
\centering
\begin{subfigure}{0.49\textwidth}
\caption{}
\includegraphics[width=\textwidth]{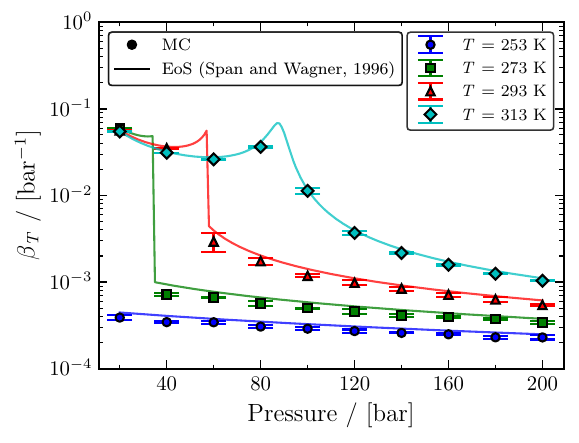}
\end{subfigure}
\begin{subfigure}{0.49\textwidth}
\caption{}
\includegraphics[width=\textwidth]{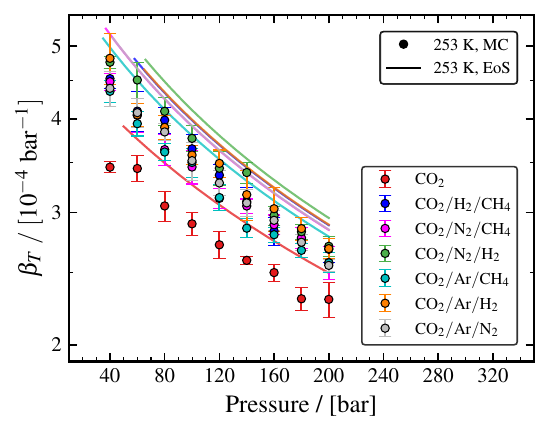}
\end{subfigure}
\begin{subfigure}{1\textwidth}
\caption{}
\includegraphics[width=\textwidth]{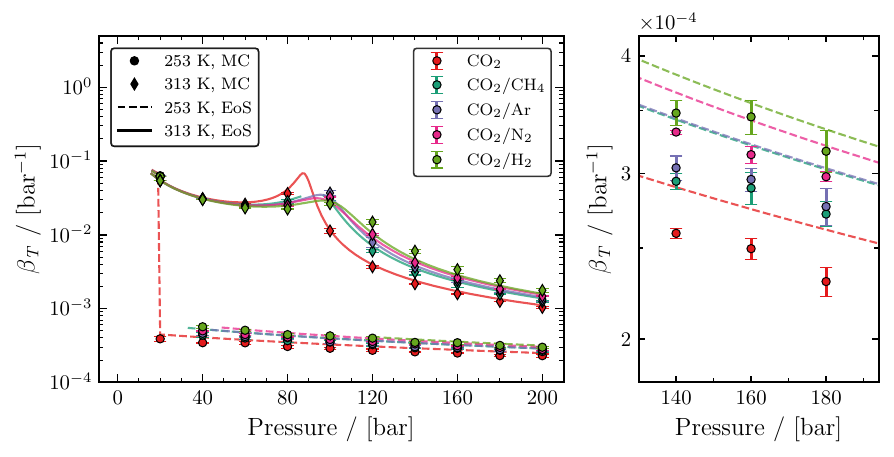}
\end{subfigure}
\caption{Computed isothermal compressibilities as a function of temperature and pressure. (a) shows the calculated isothermal compressibilities of pure \ce{CO2} from MC simulations (closed symbols) and the Span and Wagner EoS  \cite{span1996new} (solid lines) for temperatures: \SI{253}{\kelvin}, \SI{273}{\kelvin}, \SI{293}{\kelvin}, and \SI{313}{\kelvin}. (b) shows isothermal compressibilities of ternary mixtures with 96 mol\% of \ce{CO2} and 2 mol\% impurities for each of two components (\ce{CH4}, \ce{Ar}, \ce{N2}, and \ce{H2}) computed from MC simulations (closed symbols) and the GERG-2008 EoS \cite{kunz2012gerg} (lines) compared with isothermal compressibilities of pure \ce{CO2} computed from MC simulations and the Span and Wagner EoS \cite{span1996new} for temperatures: \SI{253}{\kelvin} and \SI{313}{\kelvin}. (c) shows isothermal compressibilities of binary mixtures with 95 mol\% of \ce{CO2} and 5 mol\% of impurities (\ce{CH4}, \ce{Ar}, \ce{N2}, and \ce{H2}) computed from MC simulations (closed symbols) and the GERG-2008 EoS \cite{kunz2012gerg} (lines) compared with isothermal compressibilities of pure \ce{CO2} computed from MC simulations and the Span and Wagner EoS \cite{span1996new} at \SI{253}{\kelvin}.}
\label{beta}
\end{figure}

The isothermal compressibility ($\beta_{T}$) computed from MC simulations using \cref{eqn:beta} in the \textit{NPT} ensemble for pure \ce{CO2}, binary \ce{CO2} mixtures, and ternary \ce{CO2} rich mixtures with different impurities are shown as a function of temperature and pressure in \cref{beta}. $\beta_{T}$ of pure \ce{CO2} at temperature \SI{253}{\kelvin}, \SI{293}{\kelvin}, \SI{293}{\kelvin}, and \SI{313}{\kelvin} are shown in \cref{beta} (a). The maximum value of $\beta_{T}$ was observed in the gas phase. This implies the volume change rate in response to the change in pressure is maximum when the fluid acts like an ideal gas. $\beta_{T}$ rapidly decreases with increasing pressure for pressures lower than the saturation pressure of a particular temperature. For pressures away from the saturation/Widom line, the change in the value of $\beta_{T}$ was insignificant for all temperatures, as seen in \cref{beta} (a). \cref{beta} (a) also shows that $\beta_{T}$ is dependent on temperature and $\beta_{T}$ increases with increasing temperature. $\beta_{T}$ of pure \ce{CO2} computed from MC simulations agrees qualitatively with the Span and Wagner EoS with a maximum relative deviation of ca. 19.9\% at \SI{293}{\kelvin} and 100 bar. $\beta_{T}$ of binary mixtures with 95 mol\% of \ce{CO2} and 5 mol\% of one of the impurities (\ce{CH4}, \ce{Ar}, \ce{N2}, and \ce{H2}) at temperature \SI{253}{\kelvin}, and \SI{313}{\kelvin} are shown in \cref{beta} (c). Isothermal compressibilities of binary mixtures computed from MC simulations agree fairly with the GERG-2008 EoS. The maximum relative deviation of \ce{CO2}/\ce{N2}, \ce{CO2}/\ce{CH4}, \ce{CO2}/\ce{Ar} and \ce{CO2}/\ce{H2} binary mixtures were ca. 14.1\% at \SI{253}{\kelvin} and 80 bar, ca. 17.5\% at \SI{313}{\kelvin} and 120 bar, ca. 12.9\% at \SI{253}{\kelvin} and 120 bar, and ca. 22.5\% at \SI{313}{\kelvin} and 140 bar, respectively, excluding state points close to the critical point. \cref{beta} (c) shows that the presence of impurity increases $\beta_{T}$ in the liquid and supercritical phases. In contrast, the presence of impurity decreases $\beta_{T}$ in the gas phase. Comparing $\beta_{T}$ of all binary mixtures shown in \cref{beta} (c), a binary mixture with \ce{H2} as an impurity increases the value of $\beta_{T}$ the most in the liquid and supercritical phases, followed by \ce{N2}, \ce{Ar}, and \ce{CH4}. In the gas phase, a binary mixture with \ce{H2} as an impurity decreases the value of $\beta_{T}$ followed by \ce{N2}, \ce{Ar}, and \ce{CH4}. This pattern of $\beta_{T}$ with respect to phases remained the same for both higher (10 mol\%) and lower (1 mol\%) concentrations of impurities, which is provided in Tables S73-S167 of the Supporting Information. The liquid phase $\beta_{T}$ of ternary mixtures with 96 mol\% \ce{CO2} and 2 mol\% for each of the two impurities (\ce{CH4}, \ce{Ar}, \ce{N2}, and \ce{H2}) at \SI{253}{\kelvin} are shown in \cref{beta} (b). From \cref{beta} (b), it is obvious that the presence of impurities tends to increase the value of $\beta_{T}$. Similar to thermal expansion coefficients, the effect of a particular type of impurity on $\beta_{T}$ was unclear since the change in the value of $\beta_{T}$ due to the presence of impurities was limited. However, $\beta_{T}$ computed from MC simulations for \ce{CO2} rich ternary mixtures with different impurities agrees fairly with the GERG-2008 EoS. For instance, \ce{CO2} rich ternary mixture with \ce{Ar} and \ce{CH4} as impurities results in the smallest increase in $\beta_{T}$, while \ce{CO2} rich ternary mixture with \ce{N2} and \ce{H2} as impurities lead to the largest increase in $\beta_{T}$. The difference in the value of $\beta_{T}$ due to the presence of different impurities combinations estimated from MC simulations was consistent with the GERG-2008 EoS.
 
\subsubsection{Isobaric and isochoric heat capacities}

The constant pressure and volume heat capacities are computed using \cref{eqn:CP,eqn:CV}, respectively. Residual heat capacities $C^{\mathrm{residual}}_\mathit{P}$ and $C^{\mathrm{residual}}_\mathit{V}$ are computed from MC simulations in the \textit{NPT} and \textit{NVT} ensemble respectively to obtain $c_P$ and $c_V$. \cref{fig:cp}, shows $c_P$ of pure \ce{CO2} and binary \ce{CO2} mixture and ternary \ce{CO2} mixtures as a function of temperature and pressure. $c_P$ of pure \ce{CO2} computed from MC simulations compared to the Span and Wagner EoS at temperature \SI{253}{\kelvin}, \SI{293}{\kelvin}, \SI{293}{\kelvin}, and \SI{313}{\kelvin} are shown in \cref{fig:cp} (a). Heat capacities at a constant pressure of pure \ce{CO2} are sensitive to pressures closer to the saturation/Widom line, where a sudden surge in $c_P$ is noticed. \cref{fig:cp} (a) also shows that $c_P$ increases with temperature. \textcolor{black}{The maximum value of $c_P$ observed in Fig 5 (a) at \SI{313}{\kelvin} is calculated using the Span and Wagner EoS. However, this peak value of $c_P$ cannot be predicted by MC simulations unless one considers an extremely large system. This is because the correlation length, which measures the spatial extent of spontaneous density fluctuations, diverges near the critical point \cite{frenkel2023understanding}.} $c_P$ of pure \ce{CO2} computed from MC simulations resulted in minor systematic overprediction with a maximum relative deviation of ca. 6.9\% at \SI{313}{\kelvin} and 120 bar. \textcolor{black}{At 20 bar, relative deviations of $c_P$ with Span and Wagner EoS are less than 1\% for all temperatures shown in Fig 5 (a), except at \SI{253}{\kelvin}. At \SI{253}{\kelvin} and 20 bar, $c_P$ has a relative deviation of ca. 3.8\%, which is attributed to the close distance to the vapor-liquid phase transition.} $c_P$ of binary mixtures with 95 mol\% of \ce{CO2} and 5 mol\% of one of the impurities (\ce{CH4}, \ce{Ar}, \ce{N2}, and \ce{H2}) at temperature \SI{253}{\kelvin}, and \SI{313}{\kelvin} are shown in \cref{fig:cp} (b). Similar to $c_P$ of pure \ce{CO2}, systematic overprediction of $c_P$ is observed in binary mixtures when compared to the GERG - 2008 EoS. The maximum relative deviation of \ce{CO2}/\ce{N2}, \ce{CO2}/\ce{CH4}, \ce{CO2}/\ce{Ar} and \ce{CO2}/\ce{H2} binary mixtures were ca. 8.4\% at \SI{313}{\kelvin} and 140 bar, ca. 7.3\% at \SI{313}{\kelvin} and 160 bar, ca. 6.9\% at \SI{313}{\kelvin} and 180 bar, and ca. 15.7\% at \SI{313}{\kelvin} and 140 bar, respectively, excluding state points close to the saturation/Widom line. Although $c_P$ of different binary mixtures may appear to be identical in \cref{fig:cp} (b), impurities altered the value of heat capacities significantly compared to $c_P$ of pure \ce{CO2}. The presence of impurities tends to increase the value of $c_P$ for pressure larger than the pressure at which the heat capacity peaks. Conversely, when the pressure is lower than the pressure at which the heat capacity peaks, the presence of impurities tends to decrease the value of $c_P$. For instance, $c_P$ of \ce{CO2}/\ce{H2} binary mixture at \SI{313}{\kelvin} and 160 bar were ca. 13.2\% larger than $c_P$ of pure \ce{CO2}. $c_P$ of \ce{CO2}/\ce{H2} binary mixture at \SI{313}{\kelvin} and 60 bar were ca. 13.7\% smaller than $c_P$ of pure \ce{CO2}. At conditions away from the saturation/Widom line, a particular type of impurity did not influence $c_P$ to a great extent. However, at conditions close to the saturation/Widom line, the impact of impurities on $c_P$ becomes significant based on the type of impurity where binary mixture with \ce{H2} as an impurity increase $c_P$ to a great extent followed by \ce{N2}, \ce{Ar}, and \ce{CH4}. Similar to binary mixtures with 5 mol\% impurities, binary mixtures with 1 mol\% and 10 mol\% impurities decreased and increased the value of $c_P$ compared to $c_P$ of pure \ce{CO2} before and after the maxima heat capacity, respectively, see Tables S73-S167 of the Supporting Information. $c_P$ of ternary mixtures with 96 mol\% \ce{CO2} and 2 mol\% for each of the two impurities (\ce{CH4}, \ce{Ar}, \ce{N2}, and \ce{H2}) at \SI{253}{\kelvin} are shown in \cref{fig:cp} (c). From \cref{fig:cp} (c), it is clear that the presence of impurities tends to decrease $c_P$ at \SI{253}{\kelvin}. \cref{fig:cp} (c) also confirms that the effect of a particular type of impurity combination on $c_P$ is insignificant.

Heat capacities at constant volume of pure \ce{CO2} at temperature \SI{253}{\kelvin}, \SI{293}{\kelvin}, \SI{293}{\kelvin}, and \SI{313}{\kelvin} are compared with the Span and Wagner EoS in \cref{fig:cv} (a). The computed $c_V$ is in excellent agreement with $c_V$ obtained from the Span and Wagner EoS with a maximum relative deviation of ca. 5\% at \SI{313}{\kelvin} and 120 bar, considering conditions away from the saturation/Widom line. $c_V$ values of pure \ce{CO2} vary insignificantly for pressures larger than its saturation pressure, irrespective of the temperature. Similarly, $c_V$ of binary mixtures with 95 mol\% of \ce{CO2} and 5 mol\% of one of the impurities (\ce{CH4}, \ce{Ar}, \ce{N2}, and \ce{H2}) at temperature \SI{253}{\kelvin}, and \SI{313}{\kelvin} seen in \cref{fig:cv} (b) have no significant difference in the value of $c_V$ between \SI{253}{\kelvin}, and \SI{313}{\kelvin} for pressure larger than its saturation pressure. The presence of an impurity tends to decrease $c_V$ hardly, irrespective of the type of impurity. $c_V$ of ternary mixtures with 96 mol\% \ce{CO2} and 2 mol\% for each of the two impurities (\ce{CH4}, \ce{Ar}, \ce{N2}, and \ce{H2}) at \SI{253}{\kelvin} are shown in \cref{fig:cv} (c). In the reduced axis range seen in \cref{fig:cv} (c), it is clear that the presence of impurities (in \ce{CO2}/\ce{Ar}/\ce{H2}) decreases $c_V$ maximum by ca. 3\%. It is important to mention that $c_V$ can be calculated from $c_P$ using \cref{CV-R}. However, the indirect computation of $c_V$ using \cref{CV-R} is prone to high statistical uncertainties \cite{cadena2006molecular,raabe2013molecular}, which will subsequently affect the calculations of speed of sound. Hence, constant volume heat capacities are computed by sampling \cref{eqn:CV} in \textit{NVT} ensemble with a volume of the state obtained from the \textit{NPT} simulation.

\subsubsection{Joule-Thomson coefficients \texorpdfstring{($\mu_{\mathrm{JT}}$)}{Lg}}

Joule-Thomson coefficients ($\mu_{\mathrm{JT}}$) of pure \ce{CO2}, binary \ce{CO2} mixtures, and ternary \ce{CO2} rich mixtures computed using \cref{eqn:JT} as a function of temperature and pressure are shown in \cref{mu_jt}. $\mu_{\mathrm{JT}}$ decreases as the temperature increases in the gas phase. Conversely, $\mu_{\mathrm{JT}}$ increases with temperature in the liquid and supercritical phases. The computed $\mu_{\mathrm{JT}}$ of pure \ce{CO2} at temperature \SI{253}{\kelvin}, \SI{293}{\kelvin}, \SI{293}{\kelvin}, and \SI{313}{\kelvin} seen in \cref{mu_jt} (a) agrees decently with $\mu_{\mathrm{JT}}$ obtained from the Span and Wagner EoS. The deviations of $\mu_{\mathrm{JT}}$ in the gas phase are larger when compared to the liquid and supercritical phases, and uncertainties of $\mu_{\mathrm{JT}}$ are quite significant for extremely low pressures. This is because of \cref{eqn:JT} used for computing $\mu_{\mathrm{JT}}$. Joule-Thomson coefficients are computed indirectly using \cref{eqn:JT}, and its uncertainties are computed using Eq. (S84) presented in Section S11 of the Supporting Information. The uncertainties computed using Eq. (S84) depend on the precision and accuracy of $\alpha_{P}$ and $c_{P}$. In particular, the computation of $\mu_{\mathrm{JT}}$ is highly sensitive to $\alpha_{P}$. Even an insignificant error in $\alpha_{P}$ computation might result in high uncertainties and deviations in $\mu_{\mathrm{JT}}$. A similar drawback of using \cref{eqn:JT} is also reported in Ref.\cite{agbodekhe2023assessment}. $\mu_{\mathrm{JT}}$ of binary mixtures with 95 mol\% of \ce{CO2} and 5 mol\% of one of the impurities (\ce{CH4}, \ce{Ar}, \ce{N2}, and \ce{H2}) at temperature \SI{253}{\kelvin}, and \SI{313}{\kelvin} are shown in \cref{mu_jt} (b). The value of $\mu_{\mathrm{JT}}$ decreases due to the presence of impurities in the gas phase and increases in liquid and supercritical phases. \ce{H2} has the most significant effect on the values of $\mu_{\mathrm{JT}}$, followed by \ce{N2}, \ce{Ar}, and \ce{CH4}. $\mu_{\mathrm{JT}}$ computed from MC simulations for binary mixtures rich in \ce{CO2} with 1 mol\%, 5 mol\%, and 10 mol\% of impurities (\ce{CH4}, \ce{Ar}, \ce{N2}, and \ce{H2}) at \SI{253}{\kelvin}, \SI{273}{\kelvin}, \SI{293}{\kelvin}, and \SI{313}{\kelvin} listed in Tables S73-S167 of the Supporting Information also shows the same qualitative trend with respect to impurity type. $\mu_{\mathrm{JT}}$ listed in Tables S73-S167 of the Supporting Information also indicates that the impact of impurities increases with increasing concentration of impurities. $\mu_{\mathrm{JT}}$ of ternary mixtures with 96 mol\% \ce{CO2} and 2 mol\% for each of the two impurities (\ce{CH4}, \ce{Ar}, \ce{N2}, and \ce{H2}) at \SI{253}{\kelvin} are shown in \cref{mu_jt} (c). $\mu_{\mathrm{JT}}$ of ternary mixtures computed from MC simulations at \SI{253}{\kelvin} agrees well with $\mu_{\mathrm{JT}}$ obtained from the GERG-2008 EoS, considering the uncertainty range. \cref{mu_jt} (c) also shows that the presence of impurities tends to increase $\mu_{\mathrm{JT}}$ of ternary mixtures significantly. The impact of impurities is crucial and should be considered in the initial computations, especially at conditions where the inversion of $\mu_{\mathrm{JT}}$ takes place. For instance, $\mu_{\mathrm{JT}}$ inversion occurs close to 80 bar for pure \ce{CO2} at \SI{253}{\kelvin} but for impure \ce{CO2} ternary mixtures at \SI{253}{\kelvin}, $\mu_{\mathrm{JT}}$ inversion occurs at pressure larger than 120 bar.

\subsubsection{Speed of sound \texorpdfstring{($c$)}{Lg}}

Speed of sound ($c$) of pure \ce{CO2}, binary \ce{CO2} mixtures, and ternary \ce{CO2} rich mixtures computed as a function of temperature and pressure using \cref{eqn:ss} from MC simulations are shown in \cref{ss}. Since $c$ is a function of $\rho$, $\beta_{T}$, $c_P$, and $c_V$, uncertainties associated with $c$ are computed using the Eq. (S78) derived in Section S11 of the Supporting Information. Speed of sound of pure \ce{CO2} computed at temperature \SI{253}{\kelvin}, \SI{293}{\kelvin}, \SI{293}{\kelvin}, and \SI{313}{\kelvin} results in overprediction compared to $c$ obtained using the Span and Wagner EoS. \textcolor{black}{Since $c$ depends on $\rho$, $\beta_{T}$, $c_P$, and $c_V$, small deviations in these dependent properties will overestimate $c$. The overestimation of $c$ is majorly due to the minor overestimation of $c_P$ and underestimation of $\beta_{T}$. For example, $c$ of pure \ce{CO2} is overestimated by a relative deviation of ca. 6.7\% at \SI{253}{\kelvin} and 200 bar, while the overestimation of $c_P$ is ca. 3.6\%, and the underestimation of $\beta_{T}$ is ca. 8.4\%.} The magnitude of overprediction was found to be less than 10\% for all temperatures and pressures. \cref{ss} (a), shows that $c$ decreases marginally till the pressure reaches its critical pressure and increases significantly for the pressure larger than the saturation pressure for a particular temperature. \cref{ss} (a), also shows that $c$ is temperature dependent and decreases with increasing temperature. The impact of impurities on $c$ are evaluated by comparing $c$ computed from MC simulation and the GERG-2008 EoS for binary mixtures consisting of 95 mol\% of \ce{CO2} and 5 mol\% of one of the impurities (\ce{CH4}, \ce{Ar}, \ce{N2}, and \ce{H2}) as shown in \cref{ss} (b). In the liquid and supercritical phases, the presence of \ce{H2} as an impurity is observed to have the largest impact on decreasing the value of $c$, followed by \ce{CH4}, \ce{Ar}, and \ce{N2}. In the gas phase, the presence of \ce{H2} as an impurity is observed to have the largest impact on increasing the value of $c$ followed by \ce{CH4}, \ce{Ar}, and \ce{N2}. The computed $c$ of binary mixtures rich in \ce{CO2} with 1 mol\% and 10 mol\% of impurity (\ce{CH4}, \ce{Ar}, \ce{N2}, and \ce{H2}) for temperatures \SI{253}{\kelvin}, \SI{273}{\kelvin}, \SI{293}{\kelvin}, and \SI{313}{\kelvin} from MC simulations and the GERG-2008 EoS, as listed in Tables S73-S167 of the Supporting Information shows the same behavior with respect to type of impurity. Speed of sound of ternary mixtures with 96 mol\% \ce{CO2} and 2 mol\% for each of the two impurities (\ce{CH4}, \ce{Ar}, \ce{N2}, and \ce{H2}) at \SI{253}{\kelvin} are shown in \cref{ss} (c). MC simulations overpredicted the value of $c$ compared to the GERG-2008 EoS. However, MC simulations closely predicted the decrease in the value of $c$ due to the presence of impurities. For instance, difference in the value of $c$ at 200 bar between pure \ce{CO2} and ternary \ce{CO2} mixture with \ce{N2} and \ce{H2} as impurities is ca. \SI{56}{\meter/\second} from MC simulations and ca. \SI{46}{\meter/\second} from the GERG-2008 EoS.

\subsection{Transport properties}
\subsubsection{Viscosities \texorpdfstring{$(\eta$)}{Lg}}

Viscosities ($\eta$) of pure \ce{CO2}, binary \ce{CO2} mixtures, and ternary \ce{CO2} rich mixtures computed from MD simulations as a function of temperature and pressure are shown in \cref{eta}. Viscosities of pure \ce{CO2} were computed at temperatures \SI{253}{\kelvin}, \SI{273}{\kelvin}, \SI{293}{\kelvin}, and \SI{313}{\kelvin}. The computed viscosities are compared with those calculated from the correlation of Laesecke et al. \cite{laesecke2017reference}. Our results show a good agreement with the model for all temperatures as seen in \cref{eta} (a) with a maximum relative deviation of ca. 14.6\% in the gas phase (at \SI{313}{\kelvin} and 40 bar), excluding state points close to the saturation/Widom line. $\eta$ at pressures below the saturation pressure for a temperature remains relatively constant, but for pressure far away from the saturation/Widom line, $\eta$ increases with increasing pressure at a constant temperature. $\eta$ is observed to be highly dependent on temperature compared to pressure as seen in \cref{eta} (a), where $\eta$ decreases with increasing temperature. To analyze the effect of impurities, $\eta$ of binary mixtures with 95 mol\% of \ce{CO2} and 5 mol\% of one of the impurities (\ce{CH4}, \ce{Ar}, \ce{N2}, and \ce{H2}) computed from MD simulations at \SI{253}{\kelvin} and \SI{313}{\kelvin} are compared with viscosities obtained from REFPROP in \cref{eta} (b). Computed $\eta$ of binary \ce{CO2} mixtures qualitatively correlate well with $\eta$ obtained from REFPROP. The existence of impurity in a \ce{CO2} mixture tends to reduce $\eta$ compared to $\eta$ of pure \ce{CO2}. Based on $\eta$ computed from MD simulations and REFPROP, it is clear that a binary mixture with \ce{H2} as impurity decreases $\eta$ the most in liquid and supercritical phases. The uncertainties range of computed viscosities from MD simulations made it difficult to interpret a particular type of impurity that impacts $\eta$ the most next to \ce{H2}. However, viscosities obtained from REFPROP indicate that a binary mixture with \ce{N2} as an impurity decreases $\eta$ after \ce{H2} followed by \ce{CH4}, and \ce{Ar} for pressure away from the saturation/Widom line at \SI{253}{\kelvin} and \SI{313}{\kelvin}. In addition to viscosities data of binary mixtures with 5 mol\% impurities, binary mixtures with 1 mol\% and 10 mol\% of impurities (\ce{CH4}, \ce{Ar}, \ce{N2}, and \ce{H2}) at temperatures \SI{253}{\kelvin}, \SI{273}{\kelvin}, \SI{293}{\kelvin}, and \SI{313}{\kelvin} computed from MD simulations are listed along with data obtained from REFPROP in Tables S73-S167 of the Supporting Information. To analyze the effect on $\eta$ due to the presence of a particular combination of impurities in \ce{CO2} rich mixtures, $\eta$ of ternary mixtures were computed from MD simulations and compared with $\eta$ obtained from REFPROP at \SI{253}{\kelvin}, as shown in \cref{eta} (c). $\eta$ of all ternary mixtures shown in \cref{eta} (c) has a concentration of 96 mol\% \ce{CO2} and 2 mol\% for each of the two impurities (\ce{CH4}, \ce{Ar}, \ce{N2}, and \ce{H2}). Comparing $\eta$ of ternary mixtures obtained from REFPROP, it is clear that mixtures with \ce{H2} as an impurity reduce the liquid viscosities of \ce{CO2} mixtures to a larger extent compared to mixtures without \ce{H2}. A similar qualitative trend of $\eta$ was also observed from simulations at 40, 60, and 80 bar. For higher pressures, considering the computed uncertainties, the effect of a particular combination of impurities was difficult to evaluate for a marginal decrease in the liquid viscosities. \textcolor{black}{Uncertainties associated with viscosities can be reduced by performing multiple independent simulations, but this will increase computational costs \cite{hulikal2024mutual}. We refrained from conducting additional simulations due to the marginal difference in liquid viscosities observed between ternary mixtures in \cref{eta} (c).}

In summary, we performed molecule simulations of \ce{CO2} rich mixtures with \ce{N2}, \ce{Ar}, \ce{CH4}, and \ce{H2} as impurities using force fields mentioned in Section S1 of the Supporting Information. The thermodynamic and transport properties were computed at temperatures of \SI{253}{\kelvin}, \SI{273}{\kelvin}, \SI{239}{\kelvin}, and \SI{313}{\kelvin} and pressures up to 200 bar for pure \ce{CO2}, binary mixtures rich in \ce{CO2} with 1 mol\%, 5 mol\%, and 10 mol\% impurities. The computed thermodynamic and transport properties were found to be in good agreement with EoS for pure and binary systems, except at conditions close to the saturation/Widom line. The thermodynamic and transport properties were also computed for 24 ternary and 12 quaternary \ce{CO2} rich mixtures for various concentrations of impurities listed in Tables S169 and S362 of the Supporting Information, respectively, for temperature \SI{253}{\kelvin}, \SI{273}{\kelvin}, \SI{239}{\kelvin}, and \SI{313}{\kelvin} and pressure up to 200 bar. Results of the thermodynamic and transport properties of ternary and quaternary mixtures are provided in Sections S16 and S17 of the Supporting Information, respectively. The thermodynamic and transport properties of multicomponent \ce{CO2} mixtures were compared with the GERG-2008 EoS \textcolor{black}{\cite{kunz2012gerg} and the ECS model \cite{huber2022nist}, respectively,} and showed good agreement. We show that molecular simulations are a powerful tool to compute the thermodynamic and transport properties of multicomponent mixtures seen in \ce{CO2} transportation with a smaller system of 300 molecules. These thermophysical properties will help in modeling and designing pipelines for \ce{CO2} transportation, which will be the focus of further work.

\section{Conclusions}
\label{section-5}
In this study, the effect of impurities in \ce{CO2} rich mixtures on the value of thermodynamic and transport properties such as densities, thermal expansion coefficients, isothermal compressibilities, heat capacities at constant pressure, heat capacities at constant volume, Joule-Thomson coefficients, speed of sound, and viscosities were investigated using molecular simulations. The CFCMC method was used to compute the VLE of pure components such as \ce{CO2}, \ce{CH4}, \ce{Ar}, \ce{N2}, and \ce{H2} to validate force fields used in molecular simulations. The computed VLE of pure components showed an excellent agreement with EoS \cite{span1996new,span2000reference,tegeler1999new,leachman2009fundamental,setzmann1991new}. The phase equilibria of \ce{CO2}/\ce{Ar}, \ce{CO2}/\ce{CH4}, \ce{CO2}/\ce{N2}, and \ce{CO2}/\ce{H2} binary mixtures were also computed using the CFCMC method and compared with the GERG-2008 EoS \cite{kunz2012gerg} and data from the literature \cite{coquelet2008isothermal,wei1995vapor+,davalos1976liquid,brown1989vapor,tsang1981phase}, showing a good agreement. The thermodynamic and transport properties were computed for pure \ce{CO2} and binary and ternary mixtures rich in \ce{CO2} at temperatures \SI{253}{\kelvin}, \SI{273}{\kelvin}, \SI{293}{\kelvin}, and \SI{313}{\kelvin} and for pressures ranging from 20-200 bar using MC and MD simulations. The computed thermodynamic and transport properties of pure \ce{CO2} are in excellent agreement with corresponding values obtained from the Span and Wagner EoS \cite{span1996new}. Thermodynamic and transport properties of \ce{CO2} rich binary mixtures with 1 mol\%, 5 mol\%, and 10 mol\% concentrations of non-condensable impurities such as \ce{CH4}, \ce{Ar}, \ce{N2}, and \ce{H2} computed from simulations showed a good agreement with the GERG-2008 EoS \cite{kunz2012gerg}. The computed thermodynamic and transport properties of \ce{CO2} rich ternary mixtures with various impurities were compared with the GERG-2008 EoS \cite{kunz2012gerg}, showing a good agreement. The effect of different types of impurities on a specific thermodynamic and transport property was evaluated. Our findings show that \ce{CO2} rich mixtures with impurities have low densities compared to densities of pure \ce{CO2}. The magnitude of reduction in densities of a \ce{CO2} rich mixture depends strongly on the molecular weight of impurities present a mixture. Mixtures with molecular weight lower than pure \ce{CO2} were observed to have lower densities than pure \ce{CO2}. \ce{CO2} rich mixtures containing \ce{H2} as an impurity led to the most significant decrease in the value of thermal expansion coefficients, isothermal compressibilities, heat capacities at constant pressure, and Joule-Thomson coefficients followed by \ce{N2}, \ce{Ar}, and \ce{CH4} in the gas phase. In the liquid and supercritical phases, the presence of \ce{H2} as an impurity led to the most significant increase in the value of thermal expansion coefficients, isothermal compressibilities, heat capacities at constant pressure, and Joule-Thomson coefficients followed by \ce{N2}, \ce{Ar}, and \ce{CH4}. In contrast, the presence of \ce{H2} as an impurity in \ce{CO2} rich mixture increased the value of speed of sound in the gas phase and decreased in the liquid and supercritical phases. The order of effect due to a particular impurity on thermal expansion coefficients, isothermal compressibilities, heat capacities at constant pressure, Joule-Thomson coefficients, and speed of sound correlates with the critical temperature of impurities. In our investigation of heat capacities at constant volume, we found that the presence of impurities did not have a significant impact. Finally, differences in the value of viscosities in \ce{CO2} rich mixtures due to the presence of impurities were evaluated. Our findings showed that mixtures containing \ce{H2} as impurity significantly reduced viscosities in liquid and supercritical phases.

\section*{CRediT authorship contribution statement}
\textbf{D.Raju:} Simulations, Data analysis, Writing of original and revised manuscript. \textbf{M. Ramdin:} Conceptualization, Data analysis, Co-supervision of project, proofreading of original and revised manuscript. \textbf{Thijs J.H. Vlugt:} Conceptualization, Data curation, Main supervision of the project, Writing of original and revised manuscript.

\section*{Declaration of competing interest}

The authors declare that they have no known competing financial interests or personal relationships that could have appeared to influence the work reported in this paper.

\section*{Data availability}
Data will be made available on reasonable request.

\section*{Acknowledgments}

The work presented herein is part of the ENCASE project (A European Network of Research Infrastructures for \ce{CO2} Transport and Injection). ENCASE has received funding from the European Union’s Horizon Europe Research and Innovation program under grant Number 101094664. This work was also sponsored by NWO domain Science for the use of supercomputer facilities, with financial support from the Nederlandse Organisatie voor Wetenschappelijk Onderzoek (The Netherlands Organization for Scientific Research, NWO). The authors acknowledge the use of computational resources of the DelftBlue supercomputer, provided by Delft High Performance Computing Center (https://www.tudelft.nl/dhpc).

\clearpage
\newpage

\begin{table}[tbh!]
\centering
\begin{tabular}{c|c|c|c}
\toprule
Chemical name & Chemical formula &  CAS number & Force field \\
\hline
Carbon dioxide & \ce{CO2} & 124-38-9 & TraPPE \cite{potoff2001vapor}  \\
Nitrogen & \ce{N2} & 7727-37-9 & TraPPE \cite{potoff2001vapor}  \\
Argon & \ce{Ar} & 7440-37-1 & Garc\'{i}a-P\'{e}rez \cite{garcia2008unraveling} \\
Hydrogen & \ce{H2} & 1333-74-0 & K{\"o}ster \cite{koster2018molecular} \\
Methane & \ce{CH4} & 74-82-8 &  TraPPE \cite{potoff2001vapor} \\
\bottomrule
\end{tabular}
\caption{Description of all components used in this work. Long-range tail corrections for Lennard-Jones (LJ) interactions are used for all components.}
\label{prop_table}
\end{table}

\begin{table}[tbh!]
\centering
\caption{\ce{CO2} quality standards from the National Energy Technology Laboratory (NETL) \cite{shirley_myles_2019}, Dynamis \cite{de2007towards}, International Standard Organization (ISO) \cite{simonsen_dennis}, Porthos \cite{porthos}, and Teesside \cite{brownsort20191st}. The impurity percentages with an asterisk (*) in the NETL  \cite{shirley_myles_2019} advised limits imply that the total impurity concentration should be $\leq$ 4\%. }
\label{table1}
\begin{tabular}{c|c|c|c|c|c|c}
\hline
\multicolumn{2}{c|}{Reference} & NETL \cite{shirley_myles_2019} & Dynamis \cite{de2007towards} & \begin{tabular}[c]{@{}c@{}} ISO \\ 27913:2016 \end{tabular} \cite{simonsen_dennis} & Porthos
\cite{porthos} & Teesside \cite{brownsort20191st} \\ \hline
Component & CAS number & \multicolumn{4}{c}{Concentration (in mol\%)} \\ \hline
\ce{CO2} & 124-38-9 & $\geq$ 95\% & $>$ 95.5\% & $\geq$ 95\% & $\geq $95\% & $\geq$ 95\% \\
\ce{Ar} & 7440-37-1  & 4\%* & $<$ 4\% & - & $\leq$ 0.4\% & 1\% \\
\ce{N2} & 7727-37-9  & 4\%* & $<$ 4\% & - & $\leq$ 2.4\% & 1\% \\
\ce{H2} & 1333-74-0  & 4\%* & $<$ 4\% & - & $\leq$ 0.75\% & 1\% \\
\ce{CH4}& 74-82-8  & 4\%* & $<$ 4\% & - & $\leq$ 1\% & 1\% \\
\ce{O2} & 7782-44-7 & 4\%* & $<$ 4\% & - & 40 ppm & 10 ppm \\
\ce{CO} & 630-08-0 & 35 ppm & 200 ppm & $<$ 2\% & $\leq$ 750 ppm & 0.2 \% \\
\begin{tabular}[c]{@{}c@{}}Total (\ce{Ar}, \\ \ce{N2},\ce{H2},\ce{CH4}, \\ \ce{O2}, \ce{CO})\end{tabular} & - & $\leq$ 4\% & - & $\leq$ 4\% & $\leq$ 4\% & - \\
% \ce{H2O} & 500 *ppm & 500 *ppm & $<$ 200 *ppm & 70 *ppm & 50 *ppm \\
% \ce{H2S} & 100 *ppm & 200 *ppm & $<$ 200 *ppm & 5 *ppm & 200 *ppm \\
% \ce{SOX} & 100 *ppm & - & $<$ 50 ppm & - & 100 *ppm \\
% \ce{NOX} & 100 *ppm & - & $<$ 50 ppm & $\leq$ 5 *ppm & 100 *ppm \\
\hline
\end{tabular}
\end{table}

\begin{table}[tbh!]
\centering
\begin{tabular}{c|c|c|c|c|c}
\toprule
Components & CAS number & $T^{\mathrm{REFP}}_c$/$[\mathrm{K}]$ & $T^{\mathrm{SIM}}_c$/$[\mathrm{K}]$ & $\rho^{\mathrm{REFP}}_c$/$[\mathrm{kg}$ $\mathrm{m}^{-3}]$ & $\rho^{\mathrm{SIM}}_c$/$[\mathrm{kg}$ $\mathrm{m}^{-3}]$\\
\hline
\ce{CO2} & 124-38-9 & 304.1 &306.05 & 467.6 & 466.53
\\
\ce{N2} & 7727-37-9 & 126.20 &126.30&314.40 &308.73 \\
\ce{Ar} & 7440-37-1 & 150.65 &148.93 &536 &539.70 \\
\ce{H2} & 1333-74-0 & 33.18 & 33.36 & 31.04 & 37.39 \\
\ce{CH4} & 74-82-8 & 190.6 & 191.51 & 162.1 &160.67 \\
\bottomrule
\end{tabular}
\caption{Comparison of critical temperatures ($T^{\mathrm{SIM}}_c$) and densities ($\rho^{\mathrm{SIM}}_c$) of pure components computed from simulations using the law of rectilinear diameters with critical temperatures ($T^{\mathrm{REFP}}_c$) and densities ($\rho^{\mathrm{REFP}}_c$) obtained from the REFPROP \cite{lemmon2018nist} database.}
\label{us}
\end{table}

\clearpage
\newpage

\begin{figure}[H]
\centering
\begin{subfigure}{1\textwidth}
\caption{}
\includegraphics[width=\textwidth]{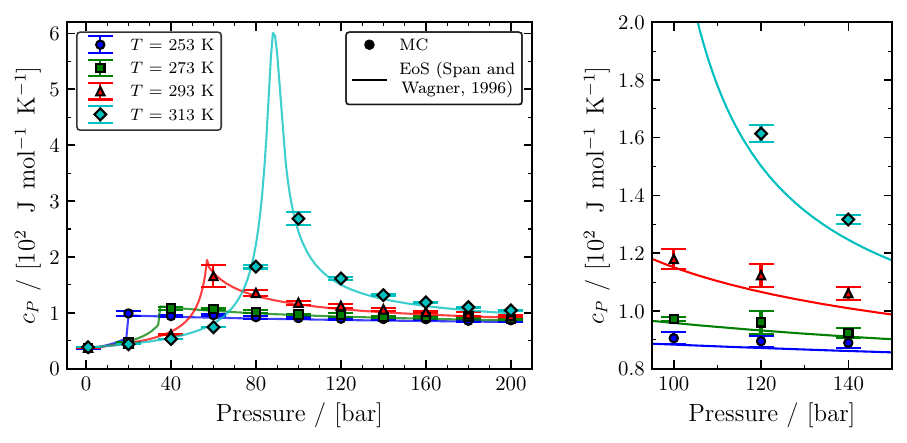}
\end{subfigure}
\begin{subfigure}{1\textwidth}
\caption{}
\includegraphics[width=\textwidth]{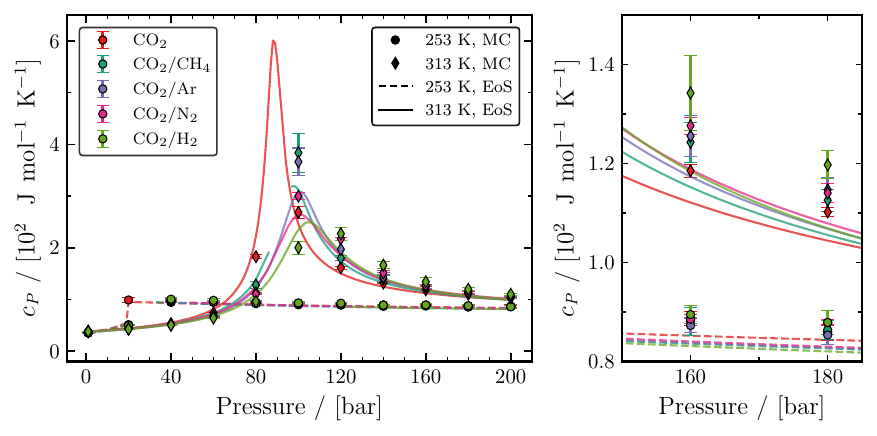}
\end{subfigure}
\caption{Computed isobaric heat capacities as a function of temperature and pressure. (a) shows the calculated isobaric heat capacities of pure \ce{CO2} from MC simulations (closed symbols) and the Span and Wagner EoS  \cite{span1996new} (solid lines) for temperatures: \SI{253}{\kelvin}, \SI{273}{\kelvin}, \SI{293}{\kelvin}, and \SI{313}{\kelvin}. (b) shows isobaric heat capacities of binary mixtures with 95 mol\% of \ce{CO2} and 5 mol\% of impurities (\ce{CH4}, \ce{Ar}, \ce{N2}, and \ce{H2}) computed from MC simulations (closed symbols) and the GERG-2008 EoS \cite{kunz2012gerg} (lines) compared with isobaric heat capacities of pure \ce{CO2} computed from MC simulations and the Span and Wagner EoS \cite{span1996new} for temperatures: \SI{253}{\kelvin} and \SI{313}{\kelvin}. (c) shows isobaric heat capacities of ternary mixtures with 96 mol\% of \ce{CO2} and 2 mol\% impurities for each of two components (\ce{CH4}, \ce{Ar}, \ce{N2}, and \ce{H2}) computed from MC simulations (closed symbols) and the GERG-2008 EoS \cite{kunz2012gerg} (lines) compared with isobaric heat capacities of pure \ce{CO2} computed from MC simulations and the Span and Wagner EoS \cite{span1996new} at \SI{253}{\kelvin}.}
\label{fig:cp}
\end{figure}

\begin{figure}[H]
\ContinuedFloat
\centering
\begin{subfigure}{0.7\textwidth}
\caption{}
\includegraphics[width=\textwidth]{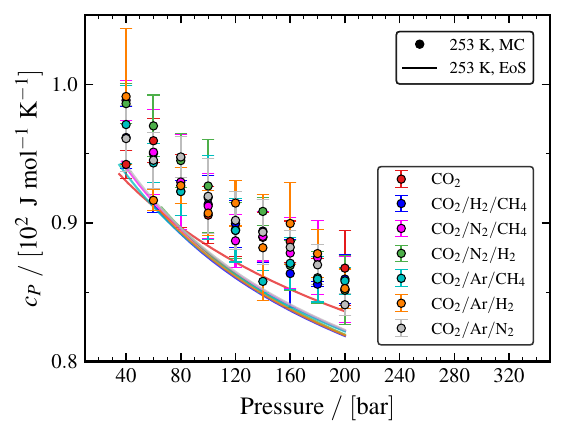}
\end{subfigure}
\caption{Computed isobaric heat capacities as a function of temperature and pressure. (a) shows the calculated isobaric heat capacities of pure \ce{CO2} from MC simulations (closed symbols) and the Span and Wagner EoS  \cite{span1996new} (solid lines) for temperatures: \SI{253}{\kelvin}, \SI{273}{\kelvin}, \SI{293}{\kelvin}, and \SI{313}{\kelvin}. (b) shows isobaric heat capacities of binary mixtures with 95 mol\% of \ce{CO2} and 5 mol\% of impurities (\ce{CH4}, \ce{Ar}, \ce{N2}, and \ce{H2}) computed from MC simulations (closed symbols) and the GERG-2008 EoS \cite{kunz2012gerg} (lines) compared with isobaric heat capacities of pure \ce{CO2} computed from MC simulations and the Span and Wagner EoS \cite{span1996new} for temperatures: \SI{253}{\kelvin} and \SI{313}{\kelvin}. (c) shows isobaric heat capacities of ternary mixtures with 96 mol\% of \ce{CO2} and 2 mol\% impurities for each of two components (\ce{CH4}, \ce{Ar}, \ce{N2}, and \ce{H2}) computed from MC simulations (closed symbols) and the GERG-2008 EoS \cite{kunz2012gerg} (lines) compared with isobaric heat capacities of pure \ce{CO2} computed from MC simulations and the Span and Wagner EoS \cite{span1996new} at \SI{253}{\kelvin}.}
\end{figure}

\begin{figure}[H]
\centering
\begin{subfigure}{0.49\textwidth}
\caption{}
\includegraphics[width=\textwidth]{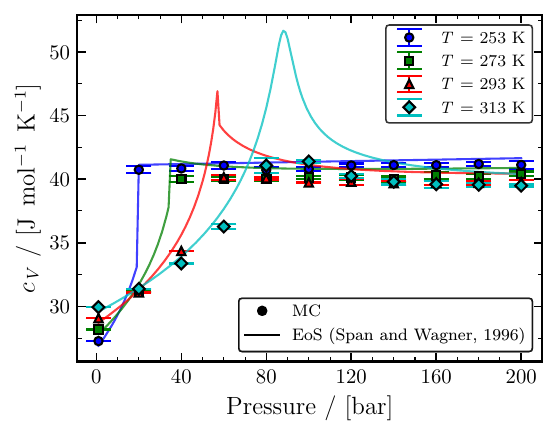}
\end{subfigure}
\begin{subfigure}{0.49\textwidth}
\caption{}
\includegraphics[width=\textwidth]{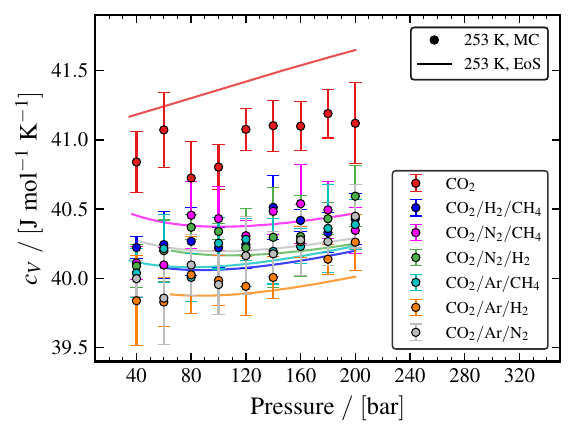}
\end{subfigure}
\begin{subfigure}{1\textwidth}
\caption{}
\includegraphics[width=\textwidth]{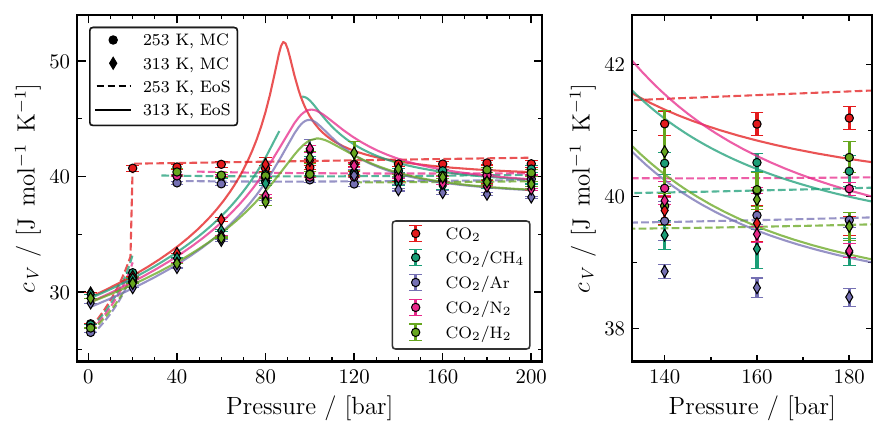}
\end{subfigure}
\caption{Computed isochoric heat capacities as a function of temperature and pressure. (a) shows the calculated isochoric heat capacities of pure \ce{CO2} from MC simulations (closed symbols) and the Span and Wagner EoS  \cite{span1996new} (solid lines) for temperatures: \SI{253}{\kelvin}, \SI{273}{\kelvin}, \SI{293}{\kelvin}, and \SI{313}{\kelvin}. (b) shows isochoric heat capacities of binary mixtures with 95 mol\% of \ce{CO2} and 5 mol\% of impurities (\ce{CH4}, \ce{Ar}, \ce{N2}, and \ce{H2}) computed from MC simulations (closed symbols) and the GERG-2008 EoS \cite{kunz2012gerg} (lines) compared with isochoric heat capacities of pure \ce{CO2} computed from MC simulations and the Span and Wagner EoS \cite{span1996new} for temperatures: \SI{253}{\kelvin} and \SI{313}{\kelvin}. (c) shows isochoric heat capacities of ternary mixtures with 96 mol\% of \ce{CO2} and 2 mol\% impurities for each of two components (\ce{CH4}, \ce{Ar}, \ce{N2}, and \ce{H2}) computed from MC simulations (closed symbols) and the GERG-2008 EoS \cite{kunz2012gerg} (lines) compared with isochoric heat capacities of pure \ce{CO2} computed from MC simulations and the Span and Wagner EoS \cite{span1996new} at \SI{253}{\kelvin}.}
\label{fig:cv}
\end{figure}

\begin{figure}[H]
\centering
\begin{subfigure}{1\textwidth}
\caption{}
\includegraphics[width=\textwidth]{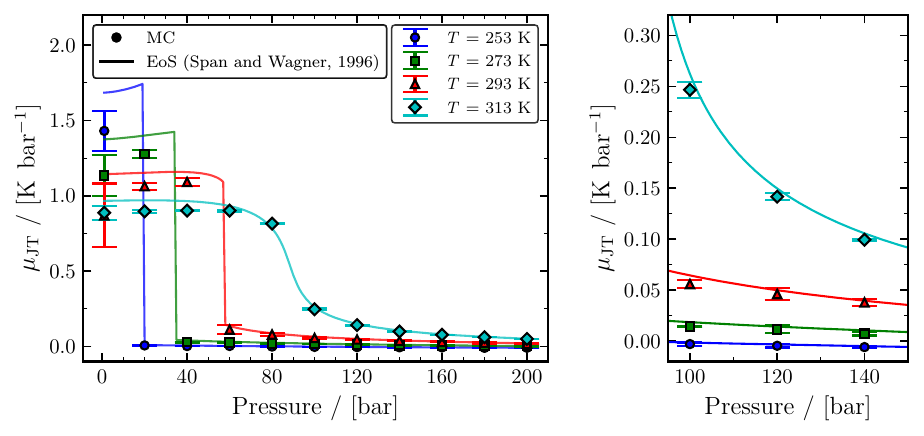}
\end{subfigure}
\begin{subfigure}{1\textwidth}
\caption{}
\includegraphics[width=\textwidth]{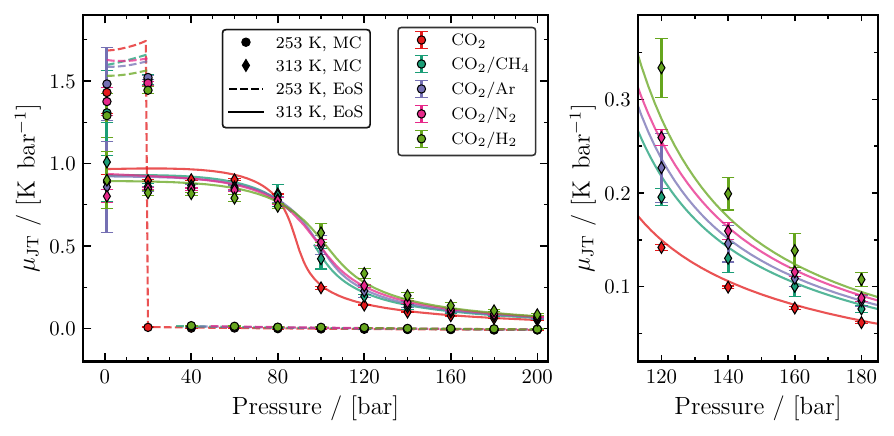}
\end{subfigure}
\caption{Computed Joule-Thomson coeﬀicients as a function of temperature and pressure. (a) shows the calculated Joule-Thomson coeﬀicients of pure \ce{CO2} from MC simulations (closed symbols) and the Span and Wagner EoS  \cite{span1996new} (solid lines) for temperatures: \SI{253}{\kelvin}, \SI{273}{\kelvin}, \SI{293}{\kelvin}, and \SI{313}{\kelvin}. (b) shows Joule-Thomson coeﬀicients of binary mixtures with 95 mol\% of \ce{CO2} and 5 mol\% of impurities (\ce{CH4}, \ce{Ar}, \ce{N2}, and \ce{H2}) computed from MC simulations (closed symbols) and the GERG-2008 EoS \cite{kunz2012gerg} (lines) compared with Joule-Thomson coeﬀicients of pure \ce{CO2} computed from MC simulations and the Span and Wagner EoS \cite{span1996new} for temperatures: \SI{253}{\kelvin} and \SI{313}{\kelvin}. (c) shows Joule-Thomson coeﬀicients of ternary mixtures with 96 mol\% of \ce{CO2} and 2 mol\% impurities for each of two components (\ce{CH4}, \ce{Ar}, \ce{N2}, and \ce{H2}) computed from MC simulations (closed symbols) and the GERG-2008 EoS \cite{kunz2012gerg} (lines) compared with Joule-Thomson coeﬀicients of pure \ce{CO2} computed from MC simulations and the Span and Wagner EoS \cite{span1996new} at \SI{253}{\kelvin}.}
\label{mu_jt}
\end{figure}

\begin{figure}[H]
\ContinuedFloat
\centering
\begin{subfigure}{0.7\textwidth}
\caption{}
\includegraphics[width=\textwidth]{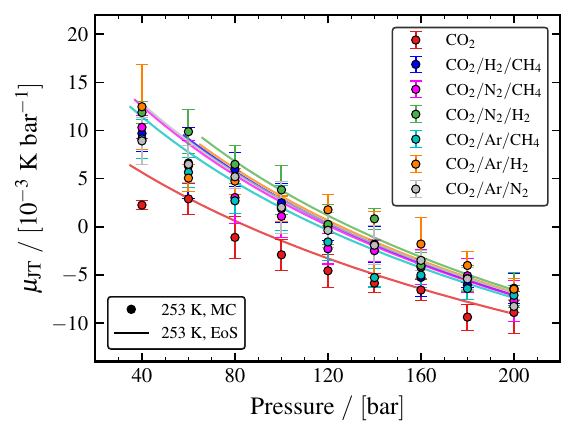}
\end{subfigure}
\caption{Computed Joule-Thomson coeﬀicients as a function of temperature and pressure. (a) shows the calculated Joule-Thomson coeﬀicients of pure \ce{CO2} from MC simulations (closed symbols) and the Span and Wagner EoS  \cite{span1996new} (solid lines) for temperatures: \SI{253}{\kelvin}, \SI{273}{\kelvin}, \SI{293}{\kelvin}, and \SI{313}{\kelvin}. (b) shows Joule-Thomson coeﬀicients of binary mixtures with 95 mol\% of \ce{CO2} and 5 mol\% of impurities (\ce{CH4}, \ce{Ar}, \ce{N2}, and \ce{H2}) computed from MC simulations (closed symbols) and the GERG-2008 EoS \cite{kunz2012gerg} (lines) compared with Joule-Thomson coeﬀicients of pure \ce{CO2} computed from MC simulations and the Span and Wagner EoS \cite{span1996new} for temperatures: \SI{253}{\kelvin} and \SI{313}{\kelvin}. (c) shows Joule-Thomson coeﬀicients of ternary mixtures with 96 mol\% of \ce{CO2} and 2 mol\% impurities for each of two components (\ce{CH4}, \ce{Ar}, \ce{N2}, and \ce{H2}) computed from MC simulations (closed symbols) and the GERG-2008 EoS \cite{kunz2012gerg} (lines) compared with Joule-Thomson coeﬀicients of pure \ce{CO2} computed from MC simulations and the Span and Wagner EoS \cite{span1996new} at \SI{253}{\kelvin}.}
\end{figure}

\begin{figure}[H]
\centering
\begin{subfigure}{0.49\textwidth}
\caption{}
\includegraphics[width=\textwidth]{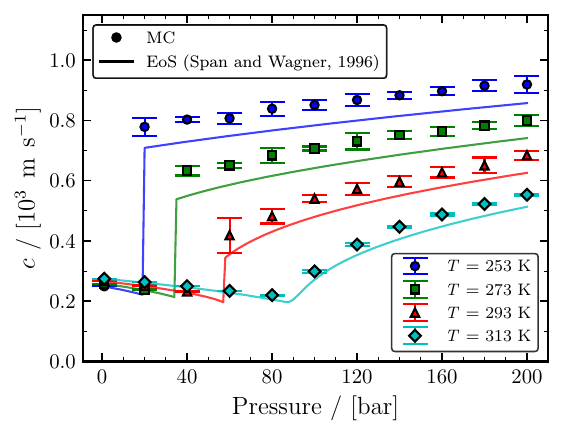}
\end{subfigure}
\begin{subfigure}{0.49\textwidth}
\caption{}
\includegraphics[width=\textwidth]{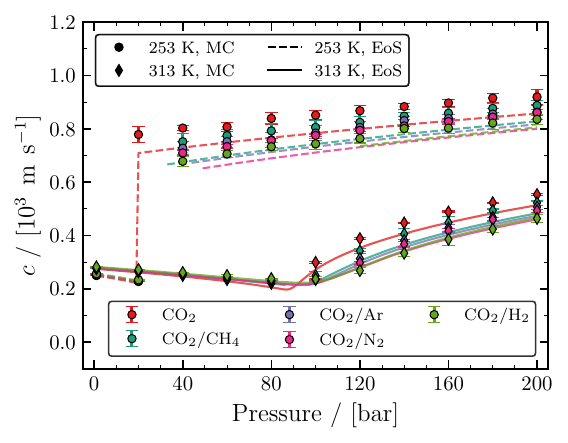}
\end{subfigure}
\begin{subfigure}{0.7\textwidth}
\caption{}
\includegraphics[width=\textwidth]{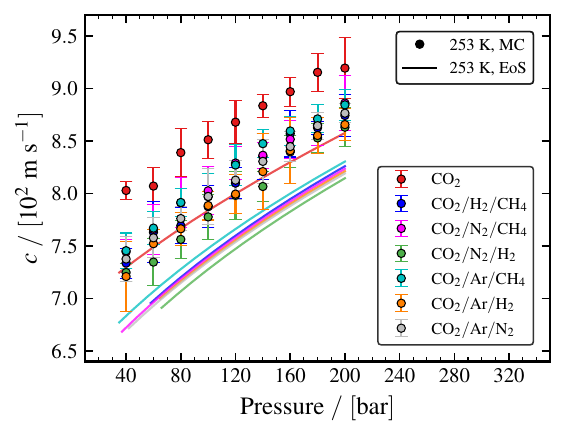}
\end{subfigure}
\caption{Computed speed of sound as a function of temperature and pressure. (a) shows the calculated speed of sound of pure \ce{CO2} from MC simulations (closed symbols) and the Span and Wagner EoS  \cite{span1996new} (solid lines) for temperatures: \SI{253}{\kelvin}, \SI{273}{\kelvin}, \SI{293}{\kelvin}, and \SI{313}{\kelvin}. (b) shows speed of sound of binary mixtures with 95 mol\% of \ce{CO2} and 5 mol\% of impurities (\ce{CH4}, \ce{Ar}, \ce{N2}, and \ce{H2}) computed from MC simulations (closed symbols) and the GERG-2008 EoS \cite{kunz2012gerg} (lines) compared with speed of sound of pure \ce{CO2} computed from MC simulations and the Span and Wagner EoS \cite{span1996new} for temperatures: \SI{253}{\kelvin} and \SI{313}{\kelvin}. (c) shows the speed of sound of ternary mixtures with 96 mol\% of \ce{CO2} and 2 mol\% impurities for each of two components (\ce{CH4}, \ce{Ar}, \ce{N2}, and \ce{H2}) computed from MC simulations (closed symbols) and the GERG-2008 EoS \cite{kunz2012gerg} (lines) compared with speed of sound of pure \ce{CO2} computed from MC simulations and the Span and Wagner EoS \cite{span1996new} at \SI{253}{\kelvin}.}
\label{ss}
\end{figure}

\begin{figure}[H]
\centering
\begin{subfigure}{0.49\textwidth}
\caption{}
\includegraphics[width=\textwidth]{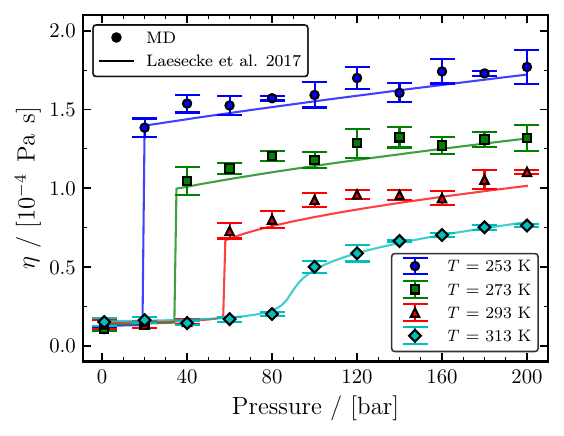}
\end{subfigure}
\begin{subfigure}{0.49\textwidth}
\caption{}
\includegraphics[width=\textwidth]{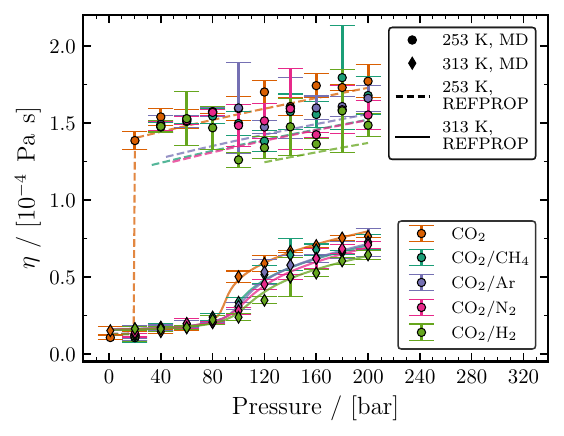}
\end{subfigure}
\begin{subfigure}{0.7\textwidth}
\caption{}
\includegraphics[width=\textwidth]{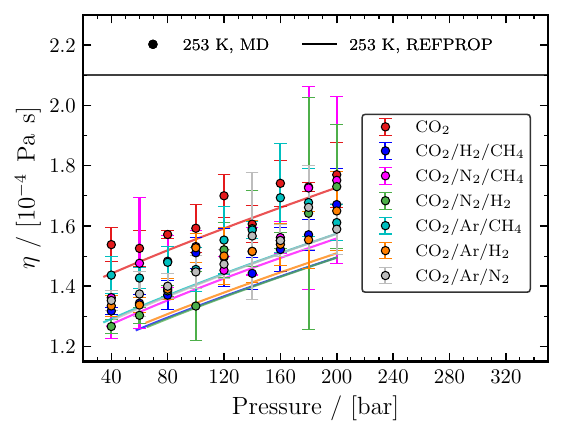}
\end{subfigure}
\caption{Computed viscosities as a function of temperature and pressure. (a) shows the calculated viscosities of pure \ce{CO2} computed from MD simulations (closed symbols) compared with the correlation of Laesecke et al. (2017) \cite{laesecke2017reference}(solid lines) for temperatures: \SI{253}{\kelvin}, \SI{273}{\kelvin}, \SI{293}{\kelvin}, and \SI{313}{\kelvin}. (b) shows viscosities of binary mixtures with 95 mol\% of \ce{CO2} and 5 mol\% of impurities (\ce{CH4}, \ce{Ar}, \ce{N2}, and \ce{H2}) computed from MD simulations (closed symbols) and REFPROP \cite{lemmon2018nist} (lines) compared with viscosities of pure \ce{CO2} computed from MD simulations and REFPROP \cite{lemmon2018nist} for temperatures: \SI{253}{\kelvin} and \SI{313}{\kelvin}. (c) shows viscosities of ternary mixtures with 96 mol\% of \ce{CO2} and 2 mol\% impurities for each of two components (\ce{CH4}, \ce{Ar}, \ce{N2}, and \ce{H2}) computed from MD simulations (closed symbols) and REFPROP \cite{lemmon2018nist} (lines) compared with viscosities of pure \ce{CO2} computed from MD simulations and REFPROP \cite{lemmon2018nist} at \SI{253}{\kelvin}.}
\label{eta}
\end{figure}

\bibliography{paper}

\end{document}